\documentclass[draft]{agujournal2019}
\usepackage{url} % this package should fix any errors with URLs in refs.
\usepackage{lineno}
\usepackage[inline]{trackchanges} % for better track changes. finalnew option will compile document with changes incorporated.
\usepackage{soul}
\nolinenumbers
\drafttrue

\journalname{Journal of Advances in Modeling Earth Systems (JAMES)}

\usepackage[super]{nth}
\usepackage{appendix}
\usepackage{siunitx}
\usepackage{booktabs}

\usepackage{amsmath}
\usepackage{amssymb}
\usepackage{amsfonts}
\usepackage{bm}

\newcommand{\grad} {\nabla}

\newcommand{\zhat} {\hat{\bm{z}}}

\newcommand{\p}[2] {\frac{\partial #1}{\partial #2}}

\newcommand*\diff{\mathop{}\!\mathrm{d}}

\newcommand{\citep}{\cite}
\newcommand{\citet}{\citeA}

\begin{document}

\title{Capturing missing physics in climate model parameterizations using neural differential equations}

\authors{
Ali Ramadhan\affil{1},
John Marshall\affil{1},
Andre Souza\affil{1},
Xin Kai Lee\affil{1,2},
Ulyana Piterbarg\affil{1},
Adeline Hillier\affil{1},
Gregory LeClaire Wagner\affil{1},
Christopher Rackauckas\affil{1},
Chris Hill\affil{1},
Jean-Michel Campin\affil{1},
Raffaele Ferrari\affil{1}
}

\affiliation{1}{Massachusetts Institute of Technology}
\affiliation{2}{Imperial College London}

\correspondingauthor{Ali Ramadhan}{alir@mit.edu}

%%%%%%%%%%%%%%%%%%%%%%%%%%%%%%%%%%
% KEY POINTS: MAX 140 CHARS EACH %
%%%%%%%%%%%%%%%%%%%%%%%%%%%%%%%%%%

\begin{keypoints}
    \item We describe a data-driven parameterization framework wherein a base parameterization is augmented with a neural network and trained online.
    \item We demonstrate that a neural differential equation can outperform existing models to parameterize vertical mixing in free convection.
    \item The approach is conservative, stable in time, independent of the time-stepper, and allows for the capture of upgradient and nonlocal fluxes.
\end{keypoints}

%%%%%%%%%%%%%%%%%%%%%%%%%%%
% ABSTRACT: MAX 250 WORDS %
%%%%%%%%%%%%%%%%%%%%%%%%%%%

\begin{abstract}
We explore how neural differential equations (NDEs) may be trained on highly resolved fluid-dynamical models of unresolved scales providing an ideal framework for data-driven parameterizations in climate models. NDEs overcome some of the limitations of traditional neural networks (NNs) in fluid dynamical applications in that they can readily incorporate conservation laws and boundary conditions and are stable when integrated over time. We advocate a method that employs a `residual' approach, in which the NN is used to improve upon an existing parameterization through the representation of residual fluxes which are not captured by the base parameterization. This reduces the amount of training required and providing a method for capturing up-gradient and nonlocal fluxes. As an illustrative example, we consider the parameterization of free convection of the oceanic boundary layer triggered by buoyancy loss at the surface. We demonstrate that a simple parameterization of the process --- convective adjustment --- can be improved upon by training a NDE against highly resolved explicit models, to capture entrainment fluxes at the base of the well-mixed layer, fluxes that convective adjustment itself cannot represent. The augmented parameterization outperforms existing commonly used parameterizations such as the K-Profile Parameterization (KPP). We showcase that the NDE performs well independent of the time-stepper and that an online training approach using differentiable simulation via the Julia scientific machine learning software stack improves accuracy by an order-of-magnitude. We conclude that NDEs provide an exciting route forward to the development of representations of sub-grid-scale processes for climate science, opening up myriad new opportunities.
\end{abstract}

%%%%%%%%%%%%%%%%%%%%%%%%%%
% SUMMARY: MAX 200 WORDS %
%%%%%%%%%%%%%%%%%%%%%%%%%%

\section*{Plain Language Summary}
Even with today’s immense computational resources, climate models cannot resolve every cloud in the atmosphere or eddying swirl in the ocean. However, collectively these small-scale turbulent processes play a key role in setting Earth’s climate. Climate models attempt to represent unresolved scales via surrogate models known as parameterizations. However, these parameterizations have limited fidelity and can exhibit structural deficiencies. Here we demonstrate that neural differential equations (NDEs) may be trained on highly resolved fluid-dynamical models of unresolved scales and act as data-driven parameterizations in an ocean model. NDEs overcome limitations of traditional neural networks in fluid dynamical applications in that they can incorporate conservation laws and are stable when integrated for long times. We argue that NDEs provide a new route forward to the development of surrogate models for climate science, opening up exciting new opportunities.

\section{Introduction: The parameterization challenge in climate modeling} \label{sec:intro}

% \begin{enumerate}
%     \item Explain why climate models need to use parameterizations.
%     \item Explain that existing parameterizations introduce biases and uncertainty in predictions.
%     \item Explain why data-driven parameterizations may help improve model fidelity.
%     \item Explain how our work relates to existing and ongoing efforts.
%     \item Emphasize that we are augmenting an existing parameterization.
% \end{enumerate}

As is often the case in science and engineering problems which address complex systems, numerical models have become central to the study of Earth's climate \citep{Hourdin17}. Climate models have historically been skillful at projecting changes in global mean surface temperature \citep{Hausfather19}. However, they often have regional deficiencies and biases which compromise future projections, especially on regional and local scales \citep{Wang14_CMIP5}. Unfortunately, more certain regional and local information is precisely what is needed to make better decisions designed to mitigate and adapt to the effects of climate change \citep{Katz13}.

A major source of uncertainty in climate \emph{projections} is due to \emph{missing physics}, that is small-scale physical processes that cannot be resolved in climate models, such as cloud formation in the atmosphere \citep{Stevens13} and small-scale boundary layer turbulence \citep{DuVivier18} and mixing by mesoscale eddies in the ocean \citep{Griffies15}. Such unresolved physical processes must be represented somehow if one is to faithfully model the evolution of Earth's climate. Instead of explicitly resolving such phenomena, which is computationally not feasible \citep{Schneider17_computing}, a more computationally efficient surrogate model, or \emph{parameterization}, is employed to represent their transport properties.

Parameterization schemes are typically developed through guidance from theory and observations but necessarily have an empirical flavor to them. For example, small-scale oceanic boundary layer turbulence cannot be resolved due to prohibitive computational costs; the coarse-grained equations are not closed and so empirical choices must be made in developing parametric representations. In the case of cloud formation, cloud microphysics and entrainment processes are not fully understood and so again, empirical choices must be made. Each parameterization thus inevitably has associated with it uncertain parameters --- such as mixing length scales or an exponent in a scaling law --- whose values must be prescribed but which are often difficult to infer from observations and associated with large uncertainties. 

Embedding multiple parameterizations into a climate model to represent myriad unresolved scales thus introduces many free parameters that must be jointly estimated. The parameters may be correlated and point estimates of the optimal parameter values may be impossible, further complicating the calibration process \citep{Souza20}. One common tuning procedure is to modify the parameters in an empirical fashion, but guided by intuition and understanding, in an effort to tune the climate model to reproduce the climate of the \nth{20} century \citep{Hourdin17} during which global observations are available. Once the model has been calibrated on historical data it is then used to extrapolate into the future. It is clear that such projections must necessarily be compromised due to the uncertainty introduced by parameterizations of unresolved scales.

Previous data-driven approaches to improving the fidelity of parameterizations have been undertaken. \citet{Souza20} attempted to automatically calibrate and quantify the uncertainty in a parameterization of oceanic convection using simulated data from high-resolution large-eddy simulations (LES). They learned that the parameterization of interest was structurally deficient because the optimal value of the parameters depended on the physical scenario. This finding suggests that some parameterizations may not be calibrated due to structural deficiencies. In such a case, developing a new parameterization would seem desirable.

\citet{Rasp18} trained a neural network to represent all atmospheric sub-grid processes in a climate model with explicit convection. Though the trained model could potentially surpass the parameterizations that it learned from in speed, it could not improve on their accuracy. Using a single model to substitute for parameterizations of multiple processes also degrades interpretibility because if the neural network behaves in unexpected ways then it is very difficult to ascertain underlying reasons. If neural networks do not exactly obey conservation laws, solutions can drift from reality ultimately leading to numerical instability or blowup. \citet{Gentine18} improved the approach by training a neural network to learn atmospheric moist convection from many superparameterized simulations that explicitly resolve moist convection. However, the 2D superparameterized models do not always faithfully resolve the inherently 3D nature of atmospheric moist convection and their neural network also does not obey conservation laws.

\citet{OGorman18} train a random forest to parameterize moist convection which has the advantage of obeying conservation laws and preserving the non-negativity of precipitation. However, random forests make predictions by matching current inputs to previously trained-upon inputs using an ensemble of decision trees. Such an approach can lead to a very large memory footprint and so be computationally demanding if trained on copious data. Furthermore, random forests do not readily generalize outside the training set. They find that training on a warm climate leads to skillful predictions for a colder climate because the extra-tropics of the warm climate provide training data for the tropics of the control climate. This reminds us of the need to train on a wide range of physical scenarios. \citet{Yuval21} extend the approach but switch to using a neural network. They train the neural network to predict coarse-grained subgrid fluxes rather than tendencies and so are able to obey conservation laws. They find that the NN performs similarly to the random forest while using much less memory.

\citet{Bolton19} train a 2D convolution neural network on subgrid eddy momentum forcings from an idealized high-resolution quasi-geostrophic ocean model mimicing a double gyre setup such as the Gulf Stream in the North Atlantic. They find it is capable of generalizing to differing viscosities and wind forcings. However, global momentum conservation must be enforced via a post-processing step to obtain optimal results. \citet{Zanna20} take a different approach and attempt to learn analytic closed-form expressions for eddy parameterizations with embedded conservation laws. The learned parameterization resembles earlier published closures but is less stable than the convolutional neural network approach. Thus, it not yet clear that equation discovery will necessarily lead to improved parameterization schemes, or that training on quasigeostrophic models can be transferred to ocean models based on more complete equation sets.

Taking a different approach to previous studies, here we describe a new route to developing data-driven parameterization in which neural differential equations (NDEs) are used to augment simple and robust existing parameterizations. The NDEs are trained on high-resolution simulations of turbulence in the surface boundary layer of the ocean and embedded into an ocean model. We describe an encouraging proof-of-concept exploration applied to free convection in the turbulent oceanic boundary layer similar in spirit to \citet{Souza20}. We outline next steps and suggest a strategy to tackle more difficult parameterization problems.

Our paper is set out as follows. In section \ref{sec:NDEs} we introduce the concept of NDEs and how they might be used to augment and improve existing parameterization schemes. In section \ref{sec:neural-free-convection} we set up an NDE which will be used to improve a simple convective adjustment parameterization of free convection. In section \ref{sec:training-data} we describe how training data can be generated from high-resolution LES simulations of the ocean boundary layer. Section \ref{sec:training} describes how the NDE is trained on this LES data using two different methods. In section \ref{sec:comparison} we compare the fidelity and performance of the trained NDE to commonly used parameterizations. In section \ref{sec:hyperparameter-optimization} we explore different neural network architectures to investigate whether nonlocality is important for the representation of free convection and to investigate the robustness of our approach to changes in neural network architecture. In section \ref{sec:conclusion} we discuss the advantages NDEs more broadly and discuss ways in which we might improve the performance and interpretability of our approach. Finally, we summarise and discuss how NDEs might be applied to more complex and challenging parameterization problems.

\section{Why use neural differential equations?} \label{sec:NDEs}

% \begin{enumerate}
%     \item Introduce neural differential equations to a climate modeling audience.
%     \item Explain the benefits of NDEs and differential versus integral control.
%     \item Explain framework for residual data-driven parameterizations.
% \end{enumerate}

\subsection{What are neural differential equations?}

Machine learning techniques have recently gained great popularity in the modeling of physical systems because of their sometimes limited ability to automatically learn nonlinear relationships from data. However, although very promising, machine learning techniques have some notable disadvantages, such as their inability to extrapolate outside the parameter range of the traning data \citep{Barnard92} and their lack of physical interpretation \citep{Fan21}. Fortunately, recent work has shown that such concerns can be overcome, or at least ameliorated, by embedding prior structural knowledge into the machine learning framework \citep{Xu20}. In the context of scientific models, the universal differential equation \citep{Rackauckas20} approach mixes known differential equation models with universal function approximators (such as neural networks) and demonstrates the ability to learn from less data and be more amenable to generalization. In this paper we will define a neural differential equation (NDE) as a differential equation where at least one term contains a neural network. The neural network contains free parameters that must be inferred for the NDE to produce useful predictions and be a useful model. Because they are phrased as a differential equation model, NDEs match the mechanistic form of physical laws, allowing for interpretability whilst presenting a learnable structure. Section \ref{sec:neural-free-convection} showcases how this NDE architecture allows for combining prior knowledge that inform existing parameterizations directly with neural networks. NDEs are trained by performing automatic differentiation through a differential equation solver provided by the Julia scientific machine learning (SciML) software stack \citep{Rackauckas17}.

% \subsection{Benefits of using neural differential equations}

% Special care has to be taken in order to allow for accurate training to occur in the presence of numerical stiffness inherent in many scientific models. For example, if one treats the upwinding semi-discretization of the advection equation $\partial_t u = \partial_x u$ using the adjoint approach of \citet{Chen18}, then one can show that the reverse solve is unconditionally unstable via Von Neumann stability analysis due to the reversal of the advection process. Thus we note that checkpointed interpolating adjoints are utilized and required for this model to prevent the gradient pathologies that are present in the training process if the ODE is reversed \citep{Kim21, Gholami19}. Specialized integrators, stabilized explicit methods of the ROCK family \citep{Abdulle02}, are utilized to efficiently handle the stiffness introduced in the PDE discretization to improve the performance and stability of the numerical solution.

\subsection{Data-driven parameterizations using a residual model}

The NDE approach has great promise, we believe, because many parameterizations of unresolved processes in climate modeling can be posed as a partial differential equation (PDE) with known terms and boundary conditions, but also unknown or incompletely known terms. Our approach is to include the known physics explicitly but to employ neural networks to represent all remaining terms. The neural network is trained on data\textemdash either observational or, as here, synthetic data from high-resolution simulations\textemdash that resolves the missing processes.

One might be tempted to use a neural network to capture all the turbulent fluxes of unresolved processes. Indeed perhaps a large enough network could be trained on enough data for this to be accomplished. However, the amount of data required may be prohibitively large and the resulting trained network might be larger than necessary. Instead, it is more efficient to harness as much physical knowledge as possible and deploy the neural network to learn physics that is difficult to parameterize. Such an approach also greatly reduces the burden on the neural network in terms of its training, which can be a great burden on computational resources.

The approach we take here, is to adopt an existing theory-driven \emph{base parameterization} that encapsulates the robust physics in a simple, interpretable manner and augment it with a neural network that predicts the unresolved physics. In this way the NDE can only improve upon the parameterization adopted.

It is important that the neural network and the NDE in to which it is embedded, is constructed in such a manner that it obeys all pertinent physical laws, such as any conservation principles and boundary conditions. For example, a base parameterization might be responsible for predicting the flux of some quantity, for example temperature $T$ which, in the application here, is the turbulent flux of heat. The neural network should not add or remove any heat from the domain, but only redistribute it within the interior. To encode this idea as a conservation law, we write the right-hand-side (RHS) of the temperature tendency equation as the divergence of a flux $\bm{F}$ thus

\begin{equation}
    \partial_t T = \grad \cdot \bm{F}
\end{equation}

enabling us to guarantee conservation by applying appropriate boundary conditions at the edges of the domain. The neural network responsible for predicting a component of the fluxes then only redistributes properties, and does not act as a source or a sink. Surface fluxes can then be prescribed as boundary conditions.

Since we know some of the robust physics that enters the RHS, but not all of the physics, we separate the flux out as

\begin{equation}
    \bm{F} = \bm{F}_\mathrm{param} + \bm{F}_\mathrm{missing}
\end{equation}

where $\bm{F}_\mathrm{param}$ will be computed using an existing base parameterization while

\begin{equation} \label{eq:F_missing}
    \bm{F}_\mathrm{missing} = \mathbb{NN}(\text{prognostic variables}, \text{surface fluxes}, \cdots)
\end{equation}

will be predicted by the neural network $\mathbb{NN}$ using available information such as the current state of the prognostic variables, surface fluxes and boundary conditions, etc. Note that  $\mathbb{NN}$, denoted such since we are using neural networks, need not be a neural network and could be any function approximator such as a Gaussian process or a polynomial series.

The guiding principles taken here are that:
\begin{itemize}
    \item $\bm{F}_\mathrm{param}$ should be \emph{simple} and based on robust well-known physics, and that
    \item the neural network is only employed to address the \emph{complicated stuff} or \emph{missing physics} we do not know how to represent.
\end{itemize}

Note that studying the fluxes predicted by the neural network could lead to an improved understanding of the missing physics and hence lead to the development of \emph{better} structured parameterizations, whether they be theory-driven or data-driven.

One major advantage of taking such a \emph{residual} parameterization approach is that an existing base parameterization can be readily used to provide an excellent first guess using, for example, down-gradient mixing assumptions. The neural network can then focus on predicting (hopefully smaller) residual fluxes, reducing the amount of training required. Moreover, the neural network can potentially capture up-gradient fluxes and nonlocal effects which are typically difficult to model using existing structured closures.

In the remainder of our paper we put these ideas in to practice in the context of free convection within an initially stratified, resting patch of ocean subject to heat loss at its upper boundary.

\section{Parameterizing free convection as a neural differential equation} \label{sec:neural-free-convection}

% \begin{enumerate}
%     \item Explain free convection and why it's important in the ocean.
%     \item Explain convective adjustment and why we use it as a base parameterization.
%     \item Derive a non-dimensional NDE for free convection.
%     \item Explain choice behind neural network architecture.
% \end{enumerate}

\subsection{Free convection}

Free convection occurs when a stably stratified fluid is cooled at its upper surface, or acted upon by some other destabilizing buoyancy flux, such as ejection of salt in ice formation. This loss of buoyancy at the surface leads to fluid parcels being made dense relative to their surroundings and so they sink, displacing buoyant fluid upwards. In this way a convectively-driven turbulent mixed layer is created that deepens with time \citep{Marshall99}. Such a process occurs everywhere in the ocean as solar radiation and clouds cause the atmosphere to warm and cool the ocean surface over the diurnal and seasonal cycles: see the review of \citep{MarshallPlumb16}. The depth of the mixed layer and the strength of the convectively-driven mixing are important factors in setting the global climate: they determine, for example, how much heat, carbon, and other soluble gases are sequestered from the atmosphere into the ocean because exchanges between the two fluids occur through this mixed layer. The surface mixed layer alone contains as much carbon as in the atmosphere. Moreover, the ocean absorbs the vast majority (over 90\%) of the excess heat in the climate system induced by greenhouse gases \citep{Resplandy19}. The mixed layer is also crucial for ocean biochemistry as it sets the vertical scale over which many marine organisms yo-yo up and down and thus the amount of sunlight they have access to and hence their growth rate. For example nutrient replenishment in the wintertime North Atlantic is associated with deep convection and to springtime blooms when the mixed layer is shallow and light is plentiful \citep[\S 7.2]{WilliamsFollows11}. Thus accurately predicting the mixed layer depth and fluxes through it is critical for ensuring the fidelity of the models used to make climate projections.

\begin{figure}[!htb]
	\centering
	\includegraphics[width=\linewidth]{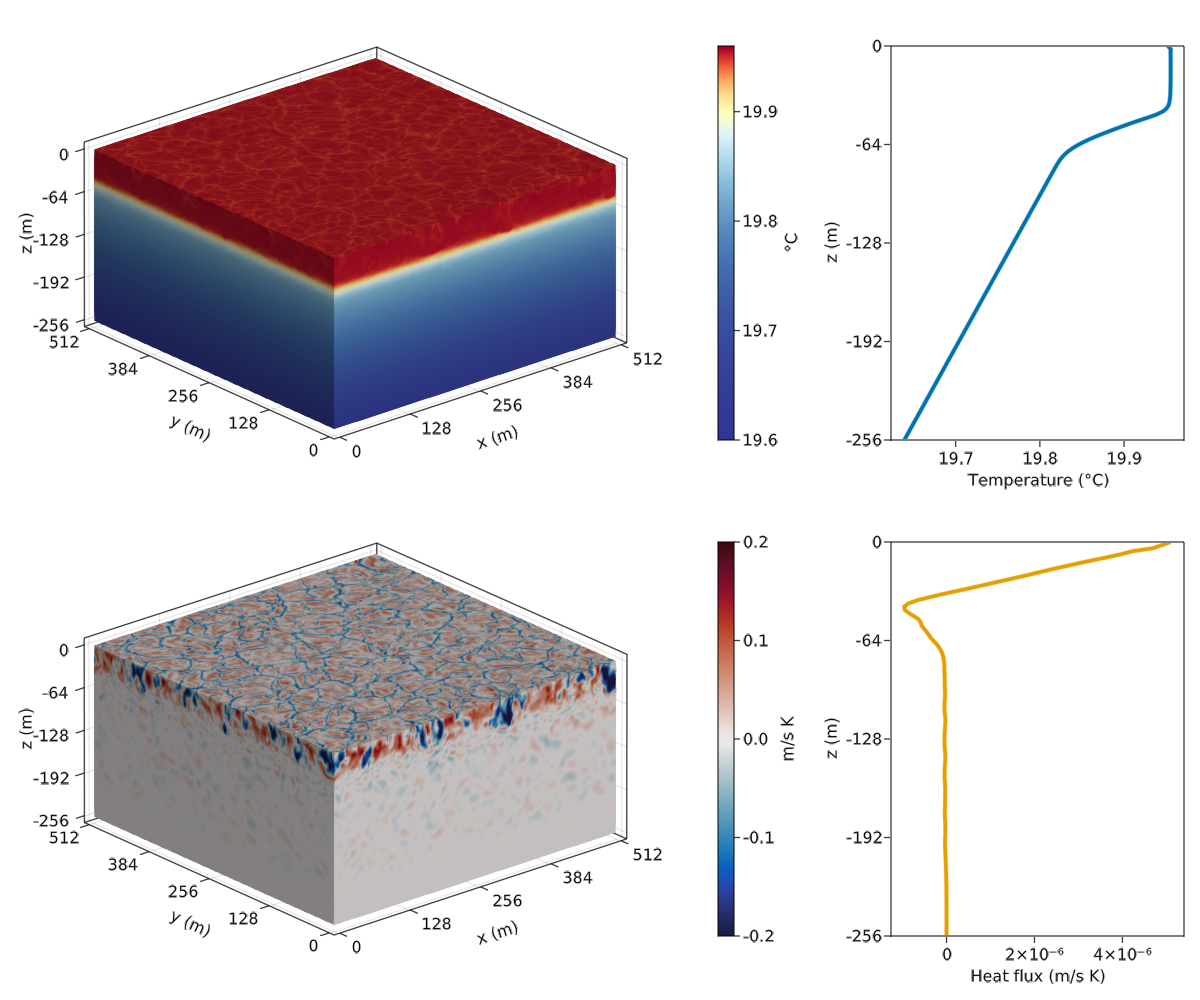}
  	\caption{Snapshot from an LES simulation of small-scale oceanic boundary layer turbulence forced by a surface cooling in a horizontally periodic domain using $256^2 \times 128$ cells and simulated using the Oceananigans.jl software package. (Left) Snapshots of the temperature field $T$ (top left) and turbulent vertical temperature flux field $w^\prime T^\prime$, proportional to the heat flux (bottom left) at $t = \SI{2}{days}$. (Right) The horizontally-averaged temperature $\overline{T}(z)$ and turbulent vertical heat flux $\overline{w^\prime T^\prime}(z)$ at the same time snapshot. Note that the heat flux profile (bottom right) includes the sub-grid scale eddy diffusivity $\overline{\kappa_e \partial_z T}(z)$ (see appendix \ref{appx:oceananigans} for details).}
 	\label{fig:oceananigans-free-convection}
\end{figure}

Wind stresses are also present at the ocean's surface which drive turbulent momentum fluxes in addition to buoyancy fluxes, further complicating the physics of boundary layer. That said, free convection is an important physical process which must be captured in models. Its accurate representation is non-trivial due to the entrainment of fluid through the base of the mixed layer where convective plumes penetrate down into the stratified fluid below. This leads to the development of an entrainment layer below the mixed layer. This entrainment physics is very difficult to represent and parameterize --- see \citet{Souza20,vanRoekel18}. Instead of attempting to develop another theory-driven parameterization of these entrainment fluxes, we will deploy a neural network within the framework of NDEs to capture them.

Figure \ref{fig:oceananigans-free-convection} shows a snapshot from a large eddy simulation of oceanic free convection. The mixed layer is the region of vertically-uniform (mixed) temperature spanning roughly the upper \SI{50}{\meter} of the domain. Over this layer the upward heat flux decreases linearly from its surface value down toward zero at the base of the layer. The entrainment region can readily be seen at the base of the mixed layer, at a depth of roughly \SI{60}{\meter}, where the heat flux becomes negative.

Explicitly simulating free convection requires solving the three-dimensional Boussinesq equations. However a look at figure \ref{fig:oceananigans-free-convection} suggests that the turbulence is horizontally homogeneous and that we may be able to predict its vertical effects without simulating the extra horizontal dimensions. A one-dimensional model of free convection in a column of water can be obtained by Reynolds averaging the advection-diffusion equation for temperature [see equation \eqref{eq:boussinesq_temperature} in appendix \ref{appx:oceananigans}] to obtain the following one-dimensional PDE for the temperature

\begin{equation} \label{eq:free-convection}
    \partial_t \overline{T} = - \partial_z \overline{w^\prime T^\prime}.
\end{equation}

Here an overline indicates a horizontal average of a three-dimensional time-varying quantity

\begin{equation}
    \overline{\phi} = \overline{\phi}(z, t) = \frac{1}{L_x L_y} \int_{0}^{L_x} \int_{0}^{L_y} \phi(x, y, z, t) \diff{x} \diff{y} 
\end{equation}

and a prime indicates a departure from the horizontal mean $\phi^\prime(x, y, z, t) = \phi(x, y, z, t) - \overline{\phi}(z, t)$: $L_x$ and $L_y$ are the domain lengths along the $x$ and $y$ dimensions: $\overline{w^\prime T^\prime}$ is the turbulent vertical heat flux responsible for redistributing heat and is the focus of our attention.

In the absence of phase changes, the fluid dynamics of the upper ocean are well described by the Boussinesq equations which can be solved numerically. Thus LES simulations can be run to provide reliable training data for $\overline{w^\prime T^\prime}$. In LES simulations the major part of the turbulent heat flux is achieved by resolved motions and an LES closure is used to represent any sub-grid fluxes through, in the present study, an eddy diffusivity. Thus equation \eqref{eq:free-convection} can be written as

\begin{equation}
    \partial_t \overline{T}
    = - \partial_z \overline{w^\prime T^\prime}
    = - \partial_z(\overline{w^\prime T^\prime}|_\text{advective} - \overline{\kappa_e \partial_z T}).
\end{equation}

Thus $\overline{w^\prime T^\prime}$ is explicitly represented via advection of temperature by the resolved flow $\overline{w^\prime T^\prime}|_\mathrm{advective}$ (computed as the resolved vertical velocity anomaly $w^\prime$ multiplied by the resolved temperature anomaly $T^\prime$) while sub-grid heat fluxes are accounted for via an eddy diffusivity $\kappa_e(x, y, z)$ modeled using an LES closure (see appendix \ref{appx:oceananigans} for more details). The diffusive eddy heat fluxes are thus a small fraction of the advective heat fluxes. In this one-dimensional model of free convection we neglect background or molecular diffusion of temperature since the molecular thermal diffusivity coefficient of seawater [$\kappa \sim \mathcal{O}(\SI{e-7}{\meter\squared\per\second})$] is orders of magnitude smaller than the typical eddy diffusivity deployed in the forward model [$\kappa_e \sim \mathcal{O}(\SI{e-2}{\meter\squared\per\second})$ in the turbulent mixed layer].

\subsection{Base parameterization: Convective adjustment}

It is well known that, in the invicid limit, if dense fluid lies above a lighter fluid, gravitational instability will ensue of the kind seen in figure \ref{fig:oceananigans-free-convection}, leading to vertical mixing \citep{Haine98}. This vertical mixing occurs due to inherently three-dimensional small-scale free convection with length scales $\mathcal{O}(\SI{e-1}{\meter})$ which cannot be resolved by global ocean models whose vertical grid spacing is $\mathcal{O}(\SI{1}{\meter})$ and horizontal grid spacing is $\mathcal{O}(\SI{10}{\kilo\meter})$ and thus the vertical mixing must be parameterized. If the mixing is not resolved numerically the model will produce statically unstable water columns with dense fluid atop lighter fluid. 

As a simple, physically-plausible base parameterization we choose convective adjustment. This will capture most of the resolved flux, but not all of it. Convective adjustment is a simple parameterization that represents this mixing via a large vertical diffusivity if vertical buoyancy gradients imply static instability \citep{Klinger96}. It can readily capture the vertical structure of the boundary layer except for the entrainment region at its base since it makes no attempt to account for entrainment processes. If a neural network can be trained to accurately predict the turbulent fluxes associated with entrainment then augmenting convective adjustment with such a neural network will likely increase the skill of the parameterization.

In this paper convective adjustment is implemented as a diffusive process in which a large diffusivity $K_\mathrm{CA}$ is turned on in regions of the water column that are statically unstable:

\begin{equation} \label{eq:convective-adjustment}
    \overline{w^\prime T^\prime}(z, t) \approx - \kappa_\mathrm{CA}(z,t) \partial_z \overline{T}(z,t)
    \quad \text{where} \quad
    \kappa_\mathrm{CA}(z,t) = 
    \begin{cases}
        K_\mathrm{CA}, & \text{if}\ \partial_z \overline{T}(z,t) < 0 \\
        0, & \text{otherwise}
    \end{cases}.
\end{equation}

% \begin{equation} \label{eq:convective_adjustment}
%     \partial_t \overline{T}(z,t) = -\partial_z \left[ \kappa_\mathrm{CA}(z,t) \partial_z \overline{T}(z,t) \right]
%     \quad \text{where} \quad
%     \kappa_\mathrm{CA}(z,t) = 
%     \begin{cases}
%         K_\mathrm{CA}, & \text{if}\ \partial_z \overline{T}(z,t) < 0 \\
%         0, & \text{otherwise}
%     \end{cases}
% \end{equation}

The one free parameter $K_\mathrm{CA}$ must be chosen which is done here by calibrating convective adjustment against the same training simulations as the NDE. We find that the optimal convective adjustment diffusivity is $K_\mathrm{CA} \approx \SI{0.2}{\meter\squared\per\second}$ (see appendix \ref{appx:calibration-ca} for details). This value is adopted and kept at the same constant value across all our experiments.

\subsection{The Neural Differential Equation for free convection}

As convective adjustment does not account for entrainment that occurs at the base of the mixed layer, we will now augment it with a neural network which will be trained to address the entrainment fluxes.

We write down the heat flux from the one-dimensional free convection model including convective adjustment thus:

\begin{equation}
    \overline{w^\prime T^\prime}
    = \overline{w^\prime T^\prime}|_\mathrm{CA} + \overline{w^\prime T^\prime}|_\mathrm{missing}
    = - \kappa_\mathrm{CA} \partial_z \overline{T} + \overline{w^\prime T^\prime}|_\mathrm{missing}.
\end{equation}

where $\overline{w^\prime T^\prime}|_\mathrm{CA}$ is the turbulent vertical heat flux predicted by convective adjustment and $\overline{w^\prime T^\prime}|_\mathrm{missing}$ is the missing portion.

Comparing the form of $\overline{w^\prime T^\prime}$ above to that obtained by horizontally averaging the equations of the LES model,

\begin{equation}
    \overline{w^\prime T^\prime} = \overline{w^\prime T^\prime}|_\mathrm{advective} - \overline{\kappa_e \partial_z T}
\end{equation}

we identify the missing heat flux as

\begin{equation} \label{eq:wT_missing}
    \overline{w^\prime T^\prime}|_\mathrm{missing} = \overline{w^\prime T^\prime}|_\mathrm{advective} - \overline{\kappa_e \partial_z T} + \kappa_\mathrm{CA} \partial_z \overline{T}.
\end{equation}

The quantity on the left-hand side, $\overline{w^\prime T^\prime}|_\mathrm{missing}$, must be learnt, while the quantities on the right are all known, being provided by the LES data (for $\overline{w^\prime T^\prime}|_\mathrm{advective} - \overline{\kappa_e \partial_z T}$) and the base parameterization ($\kappa_\mathrm{CA} \partial_z \overline{T}$). We use a neural network to represent the missing flux,

\begin{equation}
    \overline{w^\prime T^\prime}|_\mathrm{missing}(z) = \mathbb{NN}\left[\overline{T}(z)\right], 
\end{equation}

by training it to learn the relationship between equation \eqref{eq:wT_missing} and the temperature profile $\overline{T}$.

Noting that $\overline{w^\prime T^\prime} = \overline{w^\prime T^\prime}|_\mathrm{param} + \overline{w^\prime T^\prime}|_\mathrm{missing}$ and substituting the above into the PDE, equation \eqref{eq:free-convection}, we obtain our NDE --- a differential equation with a neural network representing missing physics:

\begin{equation} \label{eq:nde}
    \partial_t \overline{T}
    = - \partial_z \overline{w^\prime T^\prime}
    = - \partial_z \left[ \mathbb{NN}\left(\overline{T}\right) - \kappa_\mathrm{CA} \partial_z \overline{T} \right]
\end{equation}

The input to the NN is just the current state of the temperature profile $\overline{T}$. So the job of the NN is to predict $\overline{w^\prime T^\prime}|_\mathrm{missing}$ from the temperature profile $\overline{T}$. To train the NDE we will non-dimensionalize equation \eqref{eq:nde} (see appendix \ref{appx:nde-derivation} for a derivation of the non-dimensional NDE) used in our codes. The NDE is implemented using DifferentialEquations.jl and Julia's scientific machine learning (SciML) software stack \citep{Rackauckas17}.

\subsection{Architecture of the Neural Network}

\begin{figure}[!htb]
	\centering
	\includegraphics[width=\linewidth]{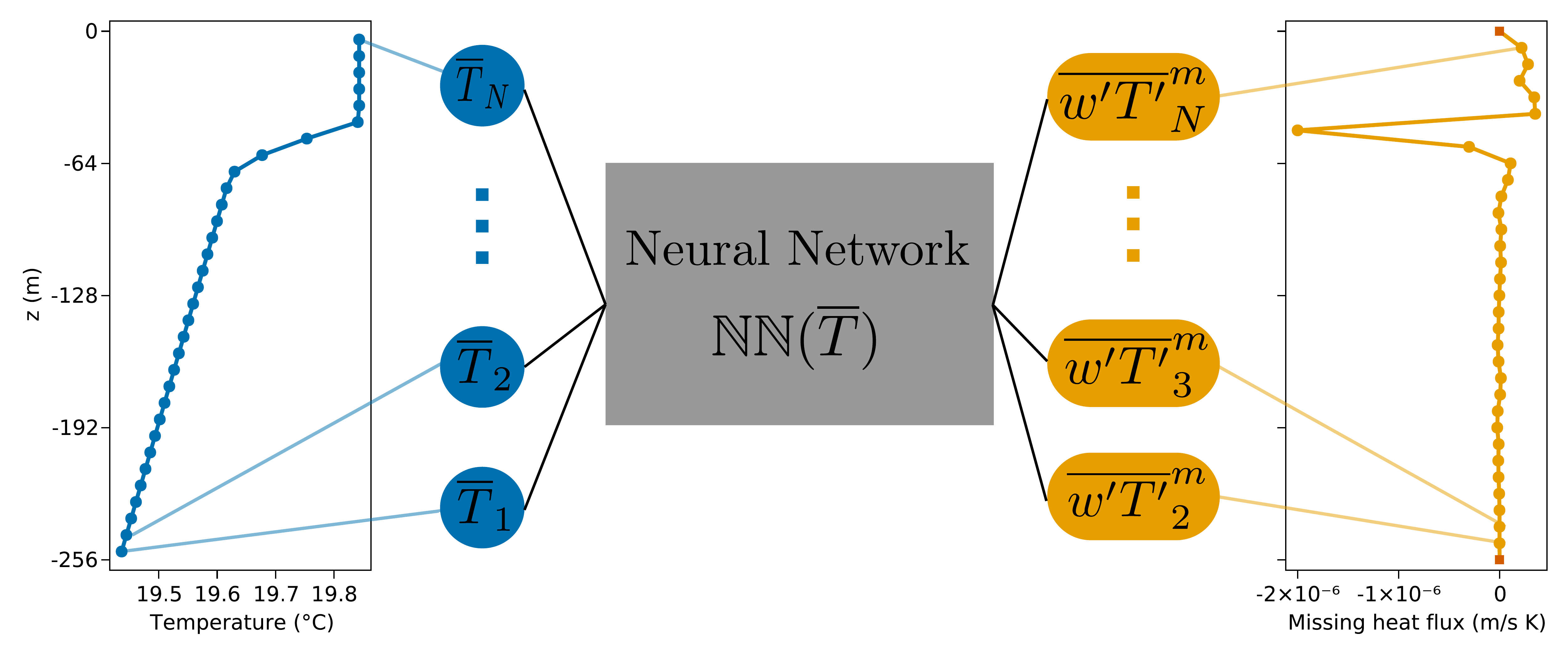}
  	\caption{Schematic representation of the architecture of the neural network. Temperature values are colored in blue while predicted missing (equivalently, residual) fluxes are colored yellow. A temperature profile consisting of $N$ values is fed into the neural network which predicts $N-1$ values for the missing heat flux in the interior of the domain. The $\overline{T}(z)$ and $\overline{w^\prime T^\prime} |_\mathrm{missing}(z)$ profiles shown here are one example from the training data. See main text for more details.}
 	\label{fig:nn-architecture}
\end{figure}

The NN predicts the missing (residual) fluxes from the prognostic variables constituting the state of the model. In the present application, as shown schematically in figure \ref{fig:nn-architecture}, we found it sufficient to supply just the temperature profile $\overline{T}(z)$ as input with $\overline{w^\prime T^\prime} |_\mathrm{missing}(z)$ as the output.

The input of the neural network is a temperature profile vector $\overline{T}(z)$ consisting of $N = 32$ floating-point values as shown on the leftmost plot of figure \ref{fig:nn-architecture} and denoted as $\overline{T}_1, \overline{T}_2, \cdots, \overline{T}_N$. The $N$ blue points show the temperature values which are the $N$ inputs to the neural network, shown as bigger blue circles connected to the values. The neural network outputs $N-1$ values predicting the missing heat flux $\overline{w^\prime T^\prime} |_\mathrm{missing}(z)$ in the interior of the domain denoted as $\overline{w^\prime T^\prime}_2^m, \overline{w^\prime T^\prime}_3^m, \cdots, \overline{w^\prime T^\prime}_N^m$. Yellow dots are fluxes predicted by the neural network while red squares are boundary fluxes that are prescribed and thus do not need to be predicted by the neural network. Knowledge of the boundary fluxes, together with the form of our NDE, clearly guarantees that conservation is obeyed. In all the calculations presented here we set $N = 32$ so that the resolution of the NDE is similar to that of a typical vertical mixing parameterization embedded in an ocean model.

We use simple feed-forward neural networks in which the input propagates through a series of layers with each layer taking an input then transforming it into an output before passing the output onto the next layer. The output may be fed through an activation function before being passed on. A layer may perform a linear transformation carried out by an activation function $\sigma$: $x \rightarrow \sigma(Wx + b)$. The layer can also perform a convolution or any other transformation on its input. The parameters of the neural network comprise entries of the weight matrix $W$ and bias vector $b$ for each layer. They are optimized using specialized optimization algorithms, many of which use various forms of gradient descent to minimize a loss function. In this way we ensure that the neural network learns some relationship between the inputs and outputs. For an introduction to neural networks and how to train them, see \citet{Mehta19}.

We choose to begin exploration by adopting a simple architecture for the neural network, denoted here by $\mathbb{NN}$, a series of fully-connected dense layers in which each layer performs the transformation $x \rightarrow \operatorname{relu}(Wx + b)$ and where the activation function is a rectified linear unit ($\operatorname{relu}$) defined as $\operatorname{relu}(x) = \operatorname{max}(0, x)$. In \S\ref{sec:hyperparameter-optimization} we will consider different $\mathbb{NN}$ architectures including wider and deeper networks as well as convolutional layers. The architectures used are catalogued in appendix \ref{appx:nn-architectures}. The Julia package Flux.jl \citep{Innes18} is used to set up the neural networks and train them.

The neural network takes $N$ input values and outputs $N-1$ values. This is because the one-dimensional model of free convection is implemented using a finite volume method on a staggered grid --- sometimes called the Arakawa C-grid after \citet{Arakawa77} --- so there are $N$ cell centers where the temperature is located but $N+1$ cell interfaces where the turbulent heat fluxes are located. The top-most and bottom-most cell interfaces correspond to the upper and lower boundaries where heat fluxes are prescribed and therefore known. This is why the neural network only predicts the remaining fluxes at the $N-1$ cell interfaces in the interior of the domain.

\section{Generation of training data using large eddy simulations} \label{sec:training-data}

% \begin{enumerate}
%     \item Explain why we use a suite of LES simulations instead of data from a big run.
%     \item Describe how we use LES to generate training data.
%     \item Describe our training and validation cases.
% \end{enumerate}

\subsection{Training data from high resolution simulations of free convection}

Ideally one would use observations of temperature profiles and turbulent heat fluxes from the real ocean to provide training data. While ocean observations exist in many regions of the ocean especially thanks to Argo floats \citep{Roemmich09}, they are not available at sufficiently high spatial and temporal resolutions in the mixed layer, and are subject to measurement error. Furthermore, observations of the contemporaneous surface fluxes are not usually available.

We could employ a very high-resolution global ocean simulation from which training data could be extracted covering a wide range of physical scenarios. However, the resolution required to resolve free convection is $\mathcal{O}(\SI{1}{\meter})$ \citep{Souza20} which is far beyond current computational capabilities in the context of a global simulation \citep{FoxKemper19}. Instead, we deploy the very high resolution LES model shown in figure \ref{fig:oceananigans-free-convection}. By varying the surface buoyancy flux and the initial stratification (by setting an initial temperature profile) the box simulations can span a large range of physical scenarios and provide a rich variety of training data in a controlled and well-defined setting.

\subsection{Simulation setup}

Simulations of free convection are run by numerically solving the Boussinesq equations with an LES closure to represent sub-grid fluxes. The LES is initialized with a stratified fluid at rest with an idealized vertical structure consisting of a weakly-stratified surface layer and a weakly-stratified abyssal/interior layer separated by a strongly-stratified thermocline. A constant surface cooling is applied via a surface buoyancy flux leading to gravitational instability and the formation of a deepening mixed layer. The LES simulations are performed using the Oceananigans.jl software package. The numerical methods employed are described in appendix \ref{appx:oceananigans} and the simulation setup is described in more detail in appendix \ref{appx:les-setup}. Horizontally-averaged output taken at regular time intervals from these simulations is used as training data. A snapshot from a typical solution is shown in figure \ref{fig:oceananigans-free-convection}.

\begin{figure}[!htb]
	\centering
	\includegraphics[width=\linewidth]{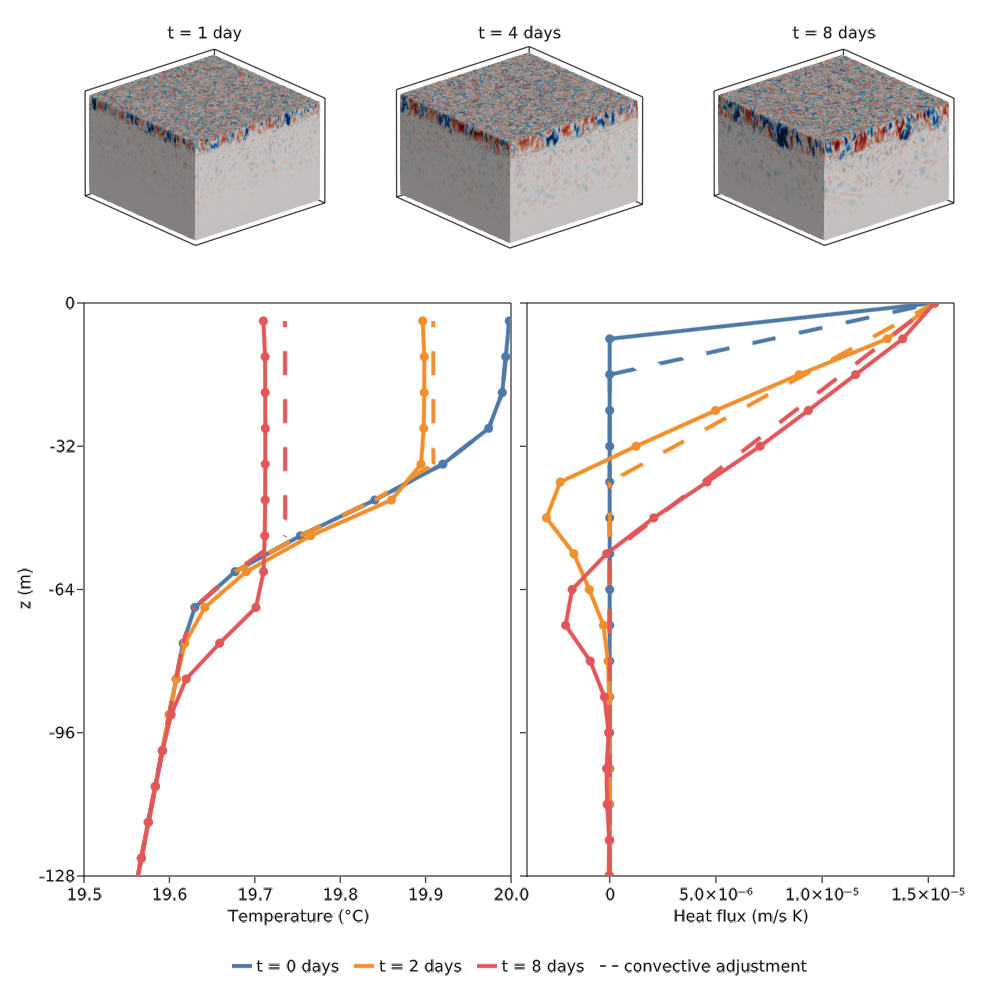}
  	\caption{Snapshots of temperature and heat flux profiles during the evolution of a convectively driven boundary layer driven by cooling from the surface. Solid lines show the true (LES) solution at different times and the points show the location of coarse-grained temperature values used by the neural network. The dashed lines show the solution produced by the convective adjustment parameterization. The heat flux profile (bottom right) includes the diffusive contribution $\overline{\kappa_e \partial_z T}(z)$ (see appendix \ref{appx:oceananigans} for more information). The profiles are taken from simulation 5 of table \ref{table:simulation-parameters} in appendix \ref{appx:simulation-parameters}. Note that only half the vertical domain is shown but the full solution extends down to $\SI{256}{\meter}$.}
 	\label{fig:training-data}
\end{figure}
% \FloatBarrier

The LES simulations are run using $256^2 \times 128$ elements with a uniform size of \SI{1}{\meter}. However, the parameterization is run in one dimension at a coarser resolution chosen, typical of that used in general circulation ocean models (GCM). This is because the parameterization is intended to eventually embed into a global GCM which typically employ coarser grids. Ocean GCMs often use a vertically stretched grid which is coarser deeper in the ocean, but reaches $\mathcal{O}(\SI{10}{\meter})$ near the surface. We use a constant vertical resolution of $\Delta z = \SI{8}{\meter}$ yielding $N_z = 32$ vertical levels. Thus the horizontally-averaged simulation output is coarse-grained from a vertical grid of 256 elements to one of 32 elements to provide training data.

The training data consists of a time series of horizontally-averaged temperature $\overline{T}(z, t)$ and horizontally-averaged turbulent vertical heat flux $\overline{w^\prime T^\prime}(z, t)$ from each simulation. They are discretized in space and time denoted, for example, by $\overline{T}_{k,n} = \overline{T}(z_k, t_n)$ where $z_k$ is the $k^\mathrm{th}$ vertical coordinate $(k \in \{1, \cdots, N_z\})$ and $t_n$ is the $n^\mathrm{th}$ time snapshot $(n \in \{1, \cdots, N_t\})$. $N_z$ is the number of vertical grid elements and $N_t$ is the number of time snapshots. Figure \ref{fig:training-data} shows a snapshot of the training data from one of the training simulations. The predictions from a convective adjustment model are also plotted using dashed lines. The difference between the LES heat flux and the heat flux from convective adjustment is precisely the missing heat flux that the neural network embedded in the NDE will learn to predict. In this way the NDE can close the gap between the simple convective adjustment parameterization scheme and the full LES simulation. With the neural network providing the missing entrainment heat fluxes, we expect the temperature profiles to evolve with skill.

\subsection{Training and validation cases}

\begin{figure}[!htb]
	\centering
	\includegraphics[width=\linewidth]{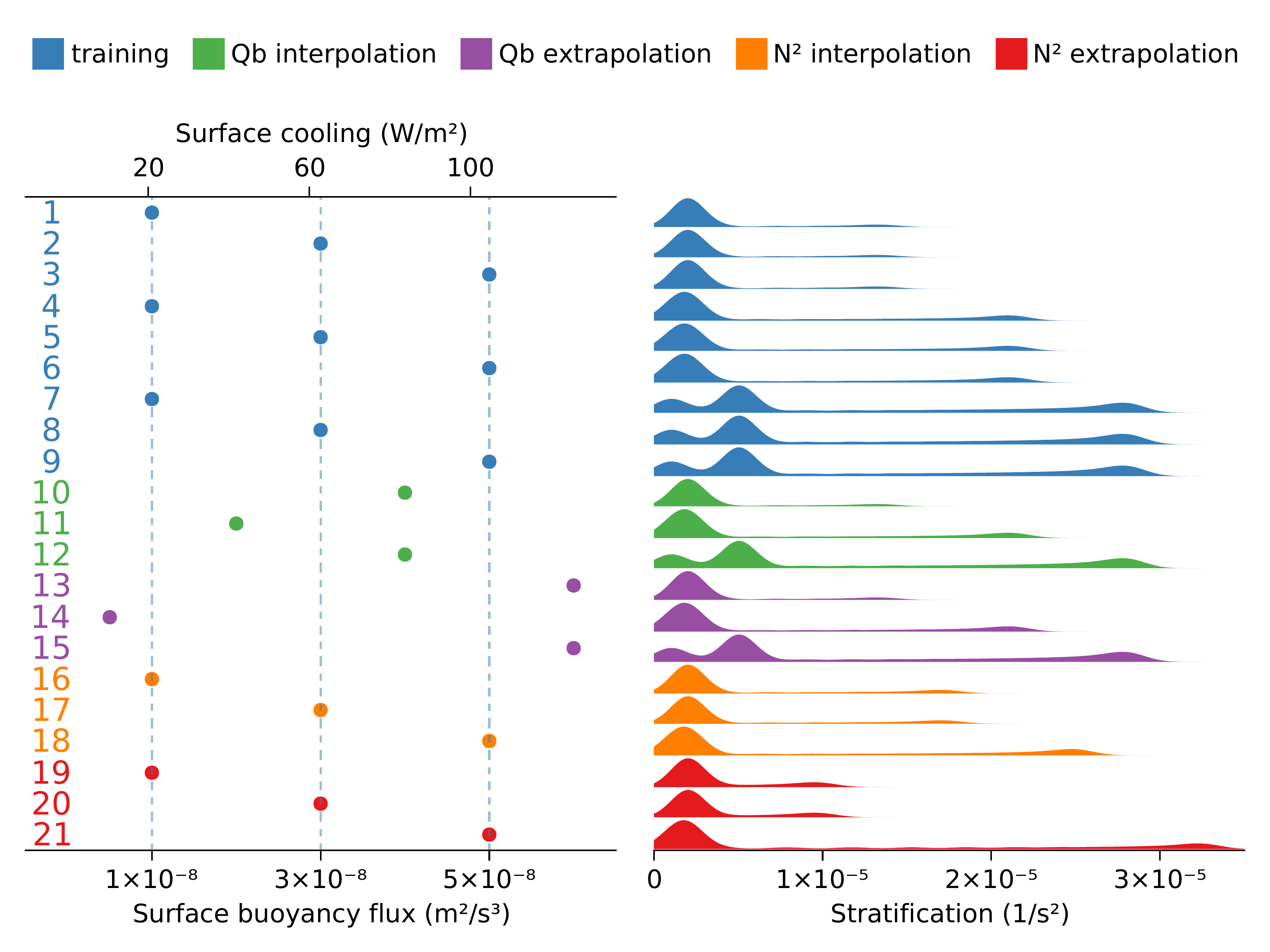}
  	\caption{Position in parameter space of simulations used for training and validation. Each simulation differs in the assumed surface heat/buoyancy flux $Q_b$ and initial stratification $N^2$ set by the initial vertical temperature distribution. Simulations 1--9 were used for training (colored in blue) while simulations 10--21 were used to test interpolation and extrapolation capabilities. (Left) The surface buoyancy flux of each simulation with dashed lines indicating the three values which provided training data. (Right) Kernel density estimates of the initial stratification distribution of each simulation.}
 	\label{fig:parameter-space}
\end{figure}

Our NDE is to be designed to perform well across a wide range of surface buoyancy fluxes and stratifications. We must necessarily train our NDE on a limited set of simulations, however, but can evaluate its ability to generalize by assessing how it performs when asked to make predictions for heat fluxes and stratifications it has not been trained on. We can test whether the NDE can interpolate --- that is make a correct prediction when the physical scenario is distinct but not more extreme than scenarios it has been trained on. We can also test whether the NDE can extrapolate --- that is perform well in scenarios outside of its training space. Interpolation and extrapolation is tested by varying both the surface buoyancy flux, $Q_b$, and the stratification, $N^2$. In the present study we used 9 training simulations and 12 testing/validation simulations split into 4 sets of 3 simulations: $Q_b$ interpolation, $Q_b$ extrapolation, $N^2$ interpolation, and $N^2$ extrapolation, as set in figure \ref{fig:parameter-space}.

% \FloatBarrier

Figure \ref{fig:parameter-space} shows the position of the 21 simulations in parameter space and whether they were used for training or validation. The simulation parameters used are tabulated in table \ref{table:simulation-parameters} in appendix \ref{appx:simulation-parameters}. The NDE was trained on 3 different values of $Q_b$ (blue points) and 3 different initial $N^2$ profiles, leading to the 9 training simulations (colored blue). To evaluate interpolation capabilities, values of $Q_b$ in between the three training values were chosen (simulations 9--12, colored green): to evaluate extrapolation capabilities, values smaller and larger than the training values were chosen (simulations 13--15, colored purple). Initial $N^2$ profiles were taken from the training simulations so that only $Q_b$ was varied.

To vary different initial $N^2$ profiles, the thicknesses and stratifications of the surface and thermocline layers were changed (see appendices \ref{appx:les-setup} and \ref{appx:simulation-parameters} for details) and the range of initial $N^2$ are visualized as kernel density estimates in figure \ref{fig:parameter-space} showing the distribution of $N^2$ present in the initial profile. To evaluate interpolation capabilities (simulations 16--18, colored orange) the simulation parameters were varied to create an initial $N^2$ profile that is different from that of the training data but which exhibits stratifications between those of the training simulations. This is evident in the plotted distributions (orange kernel density estimates) as their peaks either overlap with or are in between the peaks of the distributions for the training simulations. To evaluate extrapolation capabilities (simulations 19--21, colored red) the simulation parameters were varied to create two very weakly stratified simulations (19 and 20) and one very strongly stratified simulation (21). Values of $Q_b$ were chosen from those used in training so that only the initial $N^2$ profile was varied.

\section{Training the Neural Differential Equation} \label{sec:training}

% \begin{enumerate}
%     \item Describe optimization of base parameterization.
%     \item Describe method 1: differential control (training \emph{offline} on the fluxes).
%     \item Describe method 2: integral control (training \emph{online} on the full time series).
% \end{enumerate}

In section \ref{sec:NDEs} we discussed the key idea behind NDEs in which unknown terms are represented by neural networks and trained to learn the unknowns. This can be done in two different ways. The neural network can be trained independently of the NDE using profiles of the missing fluxes and then brought into the NDE. This first method can be called \emph{differential control} as the neural network is trained to predict fluxes at instances in time and so it learns derivatives or rates at instances in time. Alternatively, the NDE can be trained directly on the entire temperature time series with the neural network embedded within it. This second approach can be called \emph{integral control} as the neural network learns to predict the temperature integrated over time. Integral control is more computationally intensive and requires sophisticated automatic differentiation software since it involves back-propagation through the differential equation solver. In this section we will use both training methods and compare them.

Before going on it should be noted that another descriptor for our two methods could be used: \emph{differential control} is \emph{offline} training and \emph{integral control} is \emph{online} training. The distinction being that online training is performed while the parameterization is being run. Sometimes we switch between the two.

\subsection{Optimization of convective adjustment}

Before training the NDE, it is prudent to optimize our base convective adjustment parameterization. This ensures that convective adjustment performs optimally, reducing the burden on the neural network. This is done by calibrating convective adjustment against the set of training simulations described in the previous section. In the convective adjustment scheme, equation \eqref{eq:convective-adjustment}, there is a single free parameter, $K_\mathrm{CA}$. By scanning through plausible values of $K_\mathrm{CA}$ across our suite of simulations, we find the optimal value to be $K_\mathrm{CA} \approx \SI{0.2}{\meter\squared\per\second}$ (see appendix \ref{appx:calibration-ca} for details): $K_\mathrm{CA}$ was subsequently kept at a constant value of $\SI{0.2}{\meter\squared\per\second}$ in all calculations.

\subsection{Method 1: differential control --- training offline on the fluxes}

Since the neural network maps $\overline{T}(z)$ profiles to $\overline{w^\prime T^\prime} |_\mathrm{missing}(z)$ profiles (see figure \ref{fig:nn-architecture}), we can simply train the neural network to learn this relationship from appropriate training data. Training to replicate the fluxes can be thought of as a form of differential control as it is learning the instantaneous $\overline{T} \rightarrow \overline{w^\prime T^\prime} |_\mathrm{missing}(z)$ relationship. Once the neural network is trained, it can be brought into the NDE which can be immediately used to make predictions.

We seek to minimize the difference between the $\overline{w^\prime T^\prime} |_\mathrm{missing}(z)$ profile predicted by the neural network and the profile diagnosed from LES data. Thus we aim to minimize a loss function of the form

\begin{equation} \label{eq:loss-function-1}
    \mathcal{L}_1(\bm{\theta}) = \sum_{s=1}^{N_s} \mathcal{L}_{1,s}(\bm{\theta})
    \quad \mathrm{where} \quad
    \mathcal{L}_{1,s}(\bm{\theta}) =
        \frac{1}{L_z} \int_{-L_z}^0
        \left|
            \overline{w^\prime T^\prime}_s(z; \bm{\theta}) |_\mathrm{missing}^{NN}
            - \overline{w^\prime T^\prime}_s(z) |_\mathrm{missing}^{LES}
         \right|^2
         \diff{z}.
\end{equation}

Here $\mathcal{L}_1$ is the full loss function and $\mathcal{L}_{1,s}$ is the loss function over simulation $s$, $N_s = 9$ is the number of training simulations and $\bm{\theta}$ are the parameters being optimized to minimize the loss function, i.e. the neural network weights.

The neural network is trained over 5000 epochs. Each epoch is defined as one training iteration involving a full pass through the training data. The training is performed using the ADaptive Moment Estimation (ADAM) algorithm which is based on stochastic gradient descent and utilizes running averages of the gradients and its second moments \citep{Kingma14}. A learning rate (or step size) of $\eta = 10^{-3}$ is used. The top left panel of figure \ref{fig:loss-history} shows the value of the loss function \eqref{eq:loss-function-1} for the training and validation simulations as the neural network is trained (see appendix \ref{appx:nn-architectures} for training times). We see that during training the loss decreases for most of the simulation sets. The loss for the $N^2$ interpolation and extrapolation sets sharply rises then decreases before slowly increasing, possibly indicating some amount of overfitting to the training data during earlier epochs. However, the bottom left panel of figure \ref{fig:loss-history} shows that when embedded into the NDE and evaluated using a second loss function (introduced in the next subsection), training the neural network on fluxes does not actually lead to improved predictions for the temperature profile.

\subsection{Method 2: integral control --- training online on the full fluxes}

While differential control can be expected to lead to a trained neural network capable of predicting missing fluxes from the temperature profile alone, small discrepancies in the predicted fluxes may accumulate over time resulting in temperature drift. To remedy this we also try to minimize a loss function that includes the term of principal interest, the temperature profile itself.

Training to reproduce the time series can be thought of as a form of integral control as it is learning to predict behavior over the time history of the simulation, and not just at singular instants in time, and backpropagating through the differential equation solver as the NDE is simulated.

The loss function in this case takes the form

\begin{equation} \label{eq:loss-function-2}
    \mathcal{L}_2(\bm{\theta}) = \sum_{s=1}^{N_s} \mathcal{L}_{2,s}(\bm{\theta})
    \quad \mathrm{where} \quad
    \mathcal{L}_{2,s}(\bm{\theta}) =
    \frac{1}{\tau L_z} \int_0^\tau \int_{-L_z}^0
    \left|
        \overline{T}_{NDE}(z, t; \bm{\theta}) - \overline{T}_{LES}(z, t)
    \right|^2
    \diff{z} \diff{t}
\end{equation}

where $\tau$ is the length of simulation time (or window) over which $\overline{T}(z, t)$ is included in the loss function in case we do not want to train on the full time series.

For training we again find that ADAM with a learning rate of $\eta = 10^{-3}$ leads to good results. One might want to decrease $\eta$ as the loss decreases, taking smaller steps as the global minimum is approached. However, we found that incrementally or exponentially decreasing $\eta$ as a function of the epoch number did not lead to faster training or lower loss values.

Neural network weights are often initialized using schemes designed to speed up training and avoid issues of vanishing gradients during training. The neural network architectures used here are initialized in Flux.jl using Glorot uniform initialization \citep{Glorot10}. This works well with differential control training but can compromise integral control training.

Because of our formulation, the neural network is capturing a residual from a base parameterization. Therefore the starting hypothesis is that the physics parameterization is correct, which corresponds to the neural network outputting zero for everything. This is done by setting the default waits sufficiently close to zero. The initial weights are set using Glorot uniform initialization but then multiplied by $10^{-5}$. This has the effect of making the predicted missing flux negligible at epoch zero but the training process leads to the NDE slowly improving upon the base parameterization as it is trained. An alternative training method, but one we found less simple, is to not train on the full time series from epoch zero but rather to train incrementally on longer and longer time series. That is by increasing $\tau$ in equation \eqref{eq:loss-function-2}. For example, $\tau$ can be set to $\SI{3}{hours}$ for 25 epochs, then $\SI{6}{hours}$ for the next 25 epochs, then $\SI{12}{hours}$ and so on. This has the beneficial effect of slowly stabilizing the NDE as it is trained. During the integral control training process stabilized explicit time stepping methods of the ROCK family \citep{Abdulle02} are utilized to efficiently handle the stiffness introduced in the PDE discretization to improve the performance and stability of the numerical solution. We find that ROCK4 provides enough stability for all our training cases.

\begin{figure}[!htb]
	\centering
	\includegraphics[width=\linewidth]{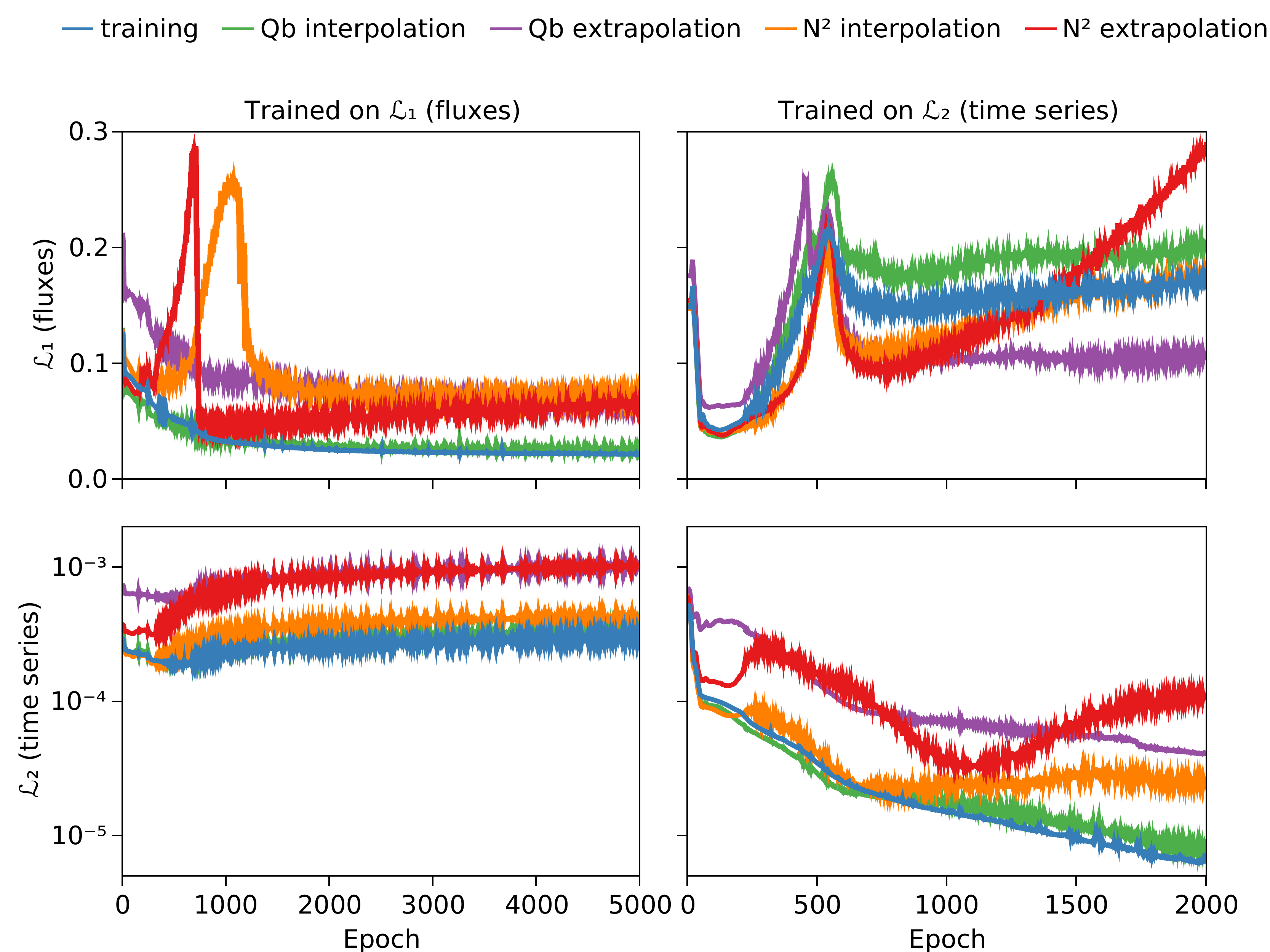}
  	\caption{Loss history of our two training methods evaluated using $\mathcal{L}_1$ [equation \eqref{eq:loss-function-1}] and $\mathcal{L}_2$ [equation \eqref{eq:loss-function-2}]. The left column shows both loss functions when training on the fluxes given by $\mathcal{L}_1$. The right column shows both loss functions when training on the time series given by $\mathcal{L}_2$. The top row shows both methods assessed using loss function $\mathcal{L}_1$ while the bottom row shows both methods assessed using $\mathcal{L}_2$. Different colors are used to indicate the training set and each testing set of simulations. The solid lines show the mean loss across all simulations in the same set.}
 	\label{fig:loss-history}
\end{figure}

Figure \ref{fig:loss-history} shows the value of both loss functions \eqref{eq:loss-function-1} and \eqref{eq:loss-function-2} for the different simulation sets. They show high-frequency noise due to the ADAM algorithm constantly attempting to ensure that the optimization does not remain in local minima. From the bottom right panel we see that augmenting convective adjustment with the trained neural network improves the loss by almost 2 orders of magnitude on the training set. The loss on the validation simulations where $Q_b$ was varied improves monotonically, perhaps indicating that overfitting is not a concern. In validation simulations where $N^2$ was varied the loss fluctuates somewhat more, potentially indicating that overfitting is occuring. Since both loss functions increase in the $N^2$ extrapolation case, overfitting is indeed likely occurring here.

Comparing the two panels on the right, we see that both loss functions drop very quickly in the first 50 epochs suggesting that training on the time series leads to learning of both the time series and the fluxes. However, as the NDE is trained more its ability to predict fluxes worsens over time. One possible explanation for this surprising result is that the neural network is being trained on instantaneous fluxes (snapshots in time) which may be noisy. Much noise is removed through horizontal averaging, however some remains. A potential remedy is to train the neural network on time-averaged horizontally-averaged fluxes.

Inspection of the two panels on the left enables us to evaluate the fidelity of the NDE trained using differential control. We see that while it performs well in predicting the fluxes (top left), the prediction of the temperature time series does not improve (bottom left). Indeed, in this case the NDE does worse at predicting the time series. This may again be associated with noise in the instantaneous fluxes and could perhaps be ameliorated somewhat by using time-averaged fluxes.

\section{Assessing the fidelity of the trained Neural Differential Equation} \label{sec:comparison}

% \begin{enumerate}
%     \item Compare the results of the NDE trained on fluxes versus on the time series.
%     \item Compare the NDE to other parameterizations (namely convective adjustment and KPP).
% \end{enumerate}

\subsection{Differential versus integral control}

In section \ref{sec:training} we described two methods of training the NDE, first via differential control and second via integral control. We now address the question of whether training an NDE on the full time series (integral control) provides any benefits over training on instantaneous flux snapshots (differential control). For the comparisons in this section the NDE is solved as part of a 1D column model simulated using Oceananigans.jl. While the NDE was trained using the explicit ROCK4 time-stepper, it performs just as well embedded in Oceananigans.jl which utilizes a split-explicit time-stepper with a second-order Adams-Bashforth explicit time-stepper and implicit Backward Euler time-stepper for convective adjustment. This demonstrates the independence of the NDE on the time stepper used operationally or during training since a neural network is only used to learn a source term for a PDE.

\begin{figure}[!htb]
	\centering
	\includegraphics[width=\linewidth]{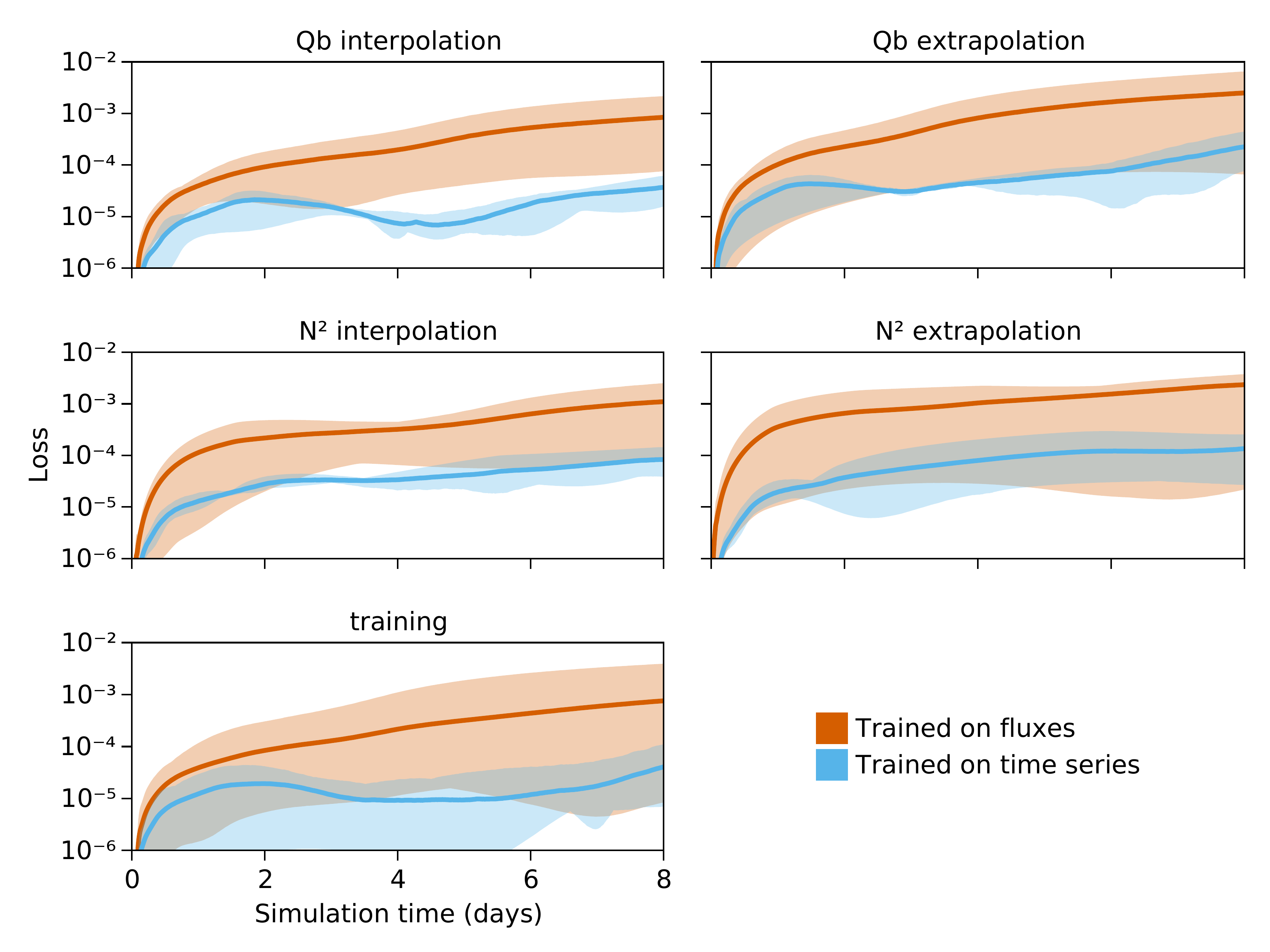}
  	\caption{Time series of the loss function $\mathcal{L}_2$ [equation \eqref{eq:loss-function-2}] over the period of mixed layer evolution. In blue is the loss of the NDE trained on the time series (integral control) while in orange is the loss of the NDE trained on the instantaneous fluxes (differential control). The solid lines show the mean loss across all simulations in the same set while the shaded area shows the minimum and maximum loss across all simulations.}
 	\label{fig:differential-vs-integral}
\end{figure}

Figure \ref{fig:differential-vs-integral} shows how our loss metric compares between our two methods over time. Using differential control the loss increases as the simulation progresses in time. This may be because, even though the neural network is trained to reproduce the fluxes at snapshots in time, errors are compounded when time-stepping forward leading to divergence. With integral control training, instead, the neural network attempts to reproduce the time series in its entirety. As a result the error does not generally grow but remains bounded. The exception is the slight increase in error near the end of the simulation period at $t = \SI{8}{days}$. The fact that the NDE solution remains congruent with the LES solution, and that by $t = \SI{8}{days}$ the loss when trained on the full time history is smaller by more than an order-of-magnitude in both training and validation sets, suggests that the use of integral control is superior to the differential control approach. This demonstrates that while loss functions can be constructed to train embedded neural networks independently of the simulation process, the approach which requires differentiating the simulation greatly improves the stability of the learned result.

\subsection{Comparison with other parameterizations} \label{ssec:comparing-parameterizations}

The main goal of developing data-driven parameterizations with NDEs is to improve upon existing parameterizations. Thus we now explore the skill of the trained NDE with existing parameterizations. In particular we compare the NDE with the base parameterization --- convective adjustment --- and a much more sophisticated scheme known as the K-Profile Parameterization (or KPP for short) described by \citet{Large94}. This is a popular vertical mixing model used by many ocean GCMs. Almost by construction, our NDE can only improve upon convective adjustment so we will focus on the comparison with KPP. The latter scheme performs well in the representation of free convection, except that it cannot easily be calibrated to work equally well for all background stratifications \citep{Souza20}. For a challenging comparison, therefore, we have also tuned KPP's parameters to perform well against our training simulations (see appendix \ref{appx:calibration-kpp} for details).

\begin{figure}[!htb]
	\centering
	\includegraphics[width=\linewidth]{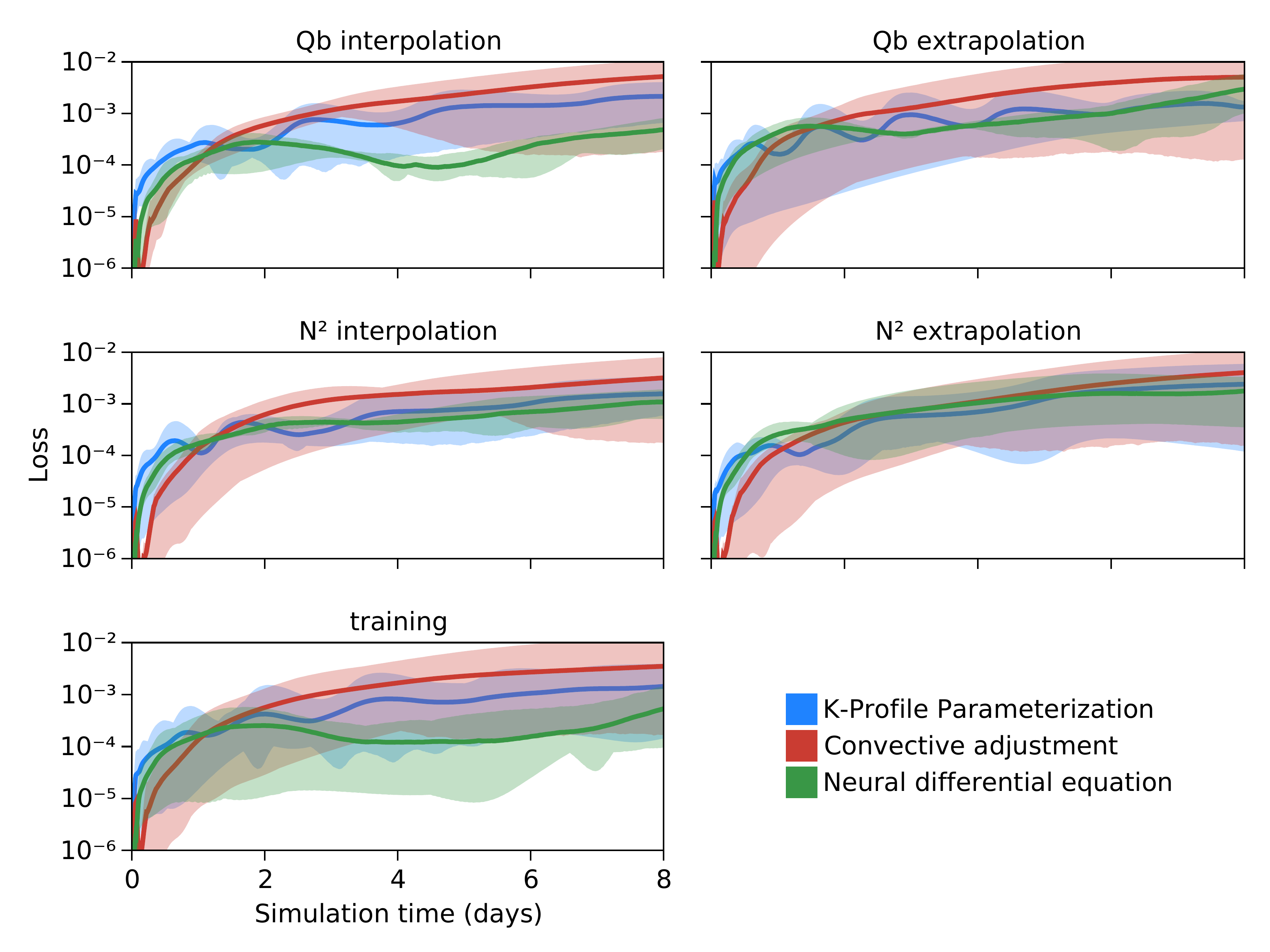}
  	\caption{The time series loss $\mathcal{L}_2$ [equation \eqref{eq:loss-function-2}] of three different parameterizations as a function of simulation time, for different sets of training and testing simulations. The solid lines show the mean loss across all simulations in the same set, while the shaded area shows the minimum and maximum loss across all simulations.}
 	\label{fig:comparing-parameterizations}
\end{figure}

Figure \ref{fig:comparing-parameterizations} shows how each parameterization performs on each of the five simulation sets. As expected, we see that in each set our NDE outperforms convective adjustment since we can only improve upon the base parameterization. As the simulation proceeds, the loss for each parameterization increases indicating that they diverge from the LES solution over time. However, our NDE generally outperforms both convective adjustment and KPP. Convective adjustment shows skill initially but deteriorates over time. For training and interpolation cases our NDE outperforms KPP while for extrapolation cases the NDE and KPP perform similarly. The NDE is capable of some extrapolation when tested against buoyancy fluxes $Q_b$ it has not seen before, but shows less skill when the stratification $N^2$ is outside the range of the training data. We will further investigate the ability to extrapolate using our NDE in section \ref{sec:hyperparameter-optimization} when we modify the architecture of the neural network.

\begin{figure}[!htb]
	\centering
	\includegraphics[width=\linewidth]{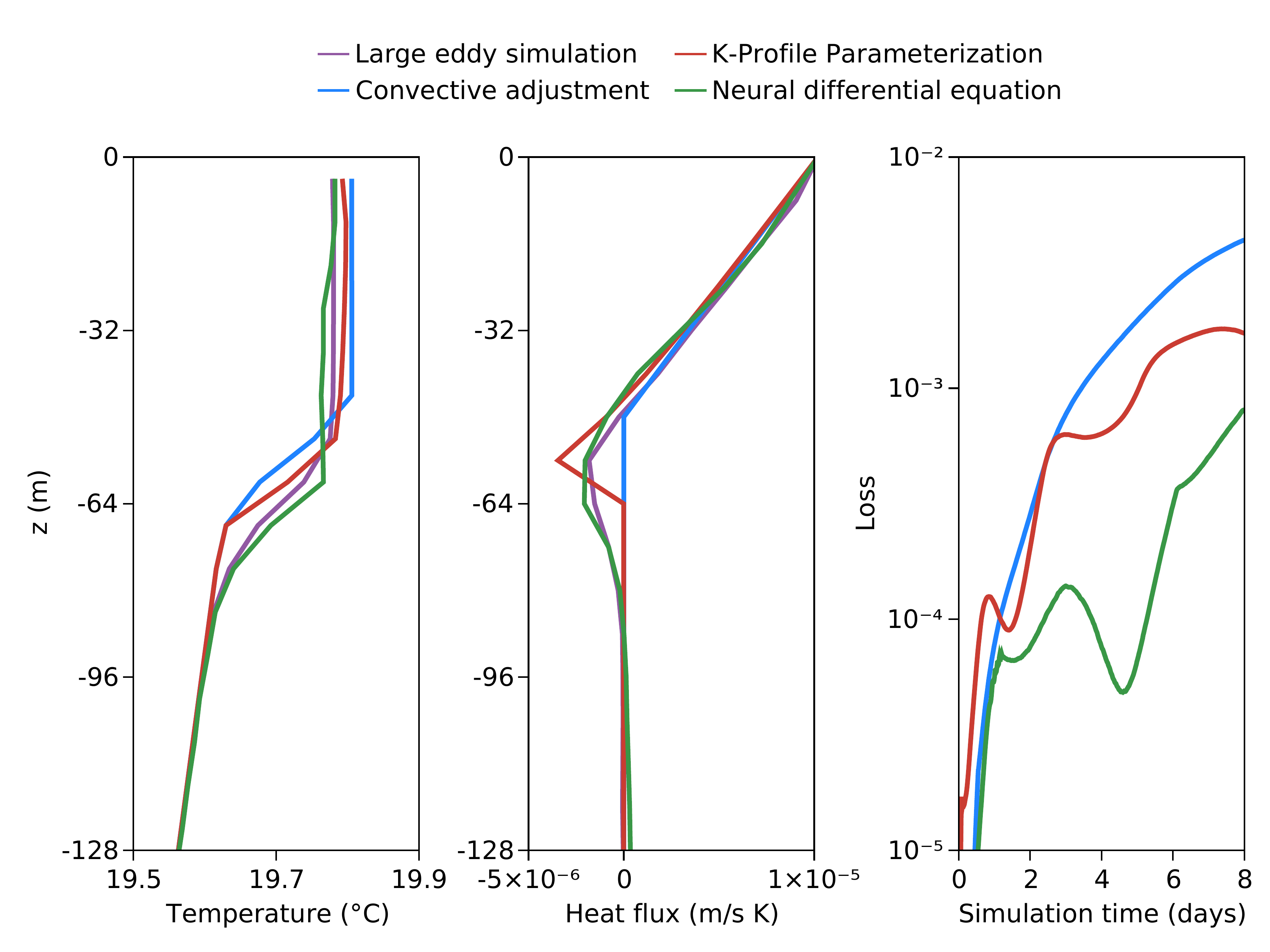}
  	\caption{Comparison of three different parameterizations against the LES (truth) solution from simulation 11 (see table \ref{table:simulation-parameters} in appendix \ref{appx:simulation-parameters}), one of our $Q_b$ interpolation simulations. (Left) The heat flux profile $\overline{w^\prime T^\prime}(z)$ predicted by the LES and by each parameterization. (Middle) The predicted temperature profiles $\overline{T}(z)$. (Right) The loss between the LES (true) solution and the predicted temperature profiles for each parameterization. Note that only half the vertical domain is shown to emphasize the structure of the mixed layer.}
 	\label{fig:comparing-parameterization-solutions}
\end{figure}

Figure \ref{fig:comparing-parameterization-solutions} updates figure \ref{fig:training-data} to include the NDE solution under integral control. The NDE solution matches the LES solution much more closely than the other parameterizations for both temperature $\overline{T}(z)$ and heat flux $\overline{w^\prime T^\prime}(z)$ profiles. Of particular interest is how closely the NDE heat flux profiles matches up with the profile from the LES simulation. The neural network was not trained on the heat flux but rather on the temperature time series. Nevertheless, it predict a physically meaningful and accurate heat flux.

\section{Hyperparameter optimization} \label{sec:hyperparameter-optimization}

Thus far we have only trained the NDE using a dense, fully-connected neural network. We will now change the architecture of the neural network in an attempt at hyperparameter optimization, i.e., optimizing the architecture of the neural network itself to improve the skill of the NDE.

There are two main reasons to optimize the architecture of the neural network: improve the fidelity of the NDE itself, and to carry out hypothesis testing using different architectures. Here we experiment with four additional architectures. For a \emph{wider} network we increase the size of the hidden layers from $4N_z$ to $8N_z$; for a \emph{deeper} network we add an extra hidden layer of size $4N_z$. We also use two neural networks with convolutional first layers and filter sizes of 2 and 4. The use of the dense and ``convolutional'' network architectures can be thought of making the assumption that the parameterization is fully nonlocal in the dense case and local in the convolutional case with the filter size determining the degree of locality. Appendix \ref{appx:nn-architectures} details the architectures used including the number of free parameters (or weights) in each one.

\begin{figure}[!htb]
	\centering
	\includegraphics[width=\linewidth]{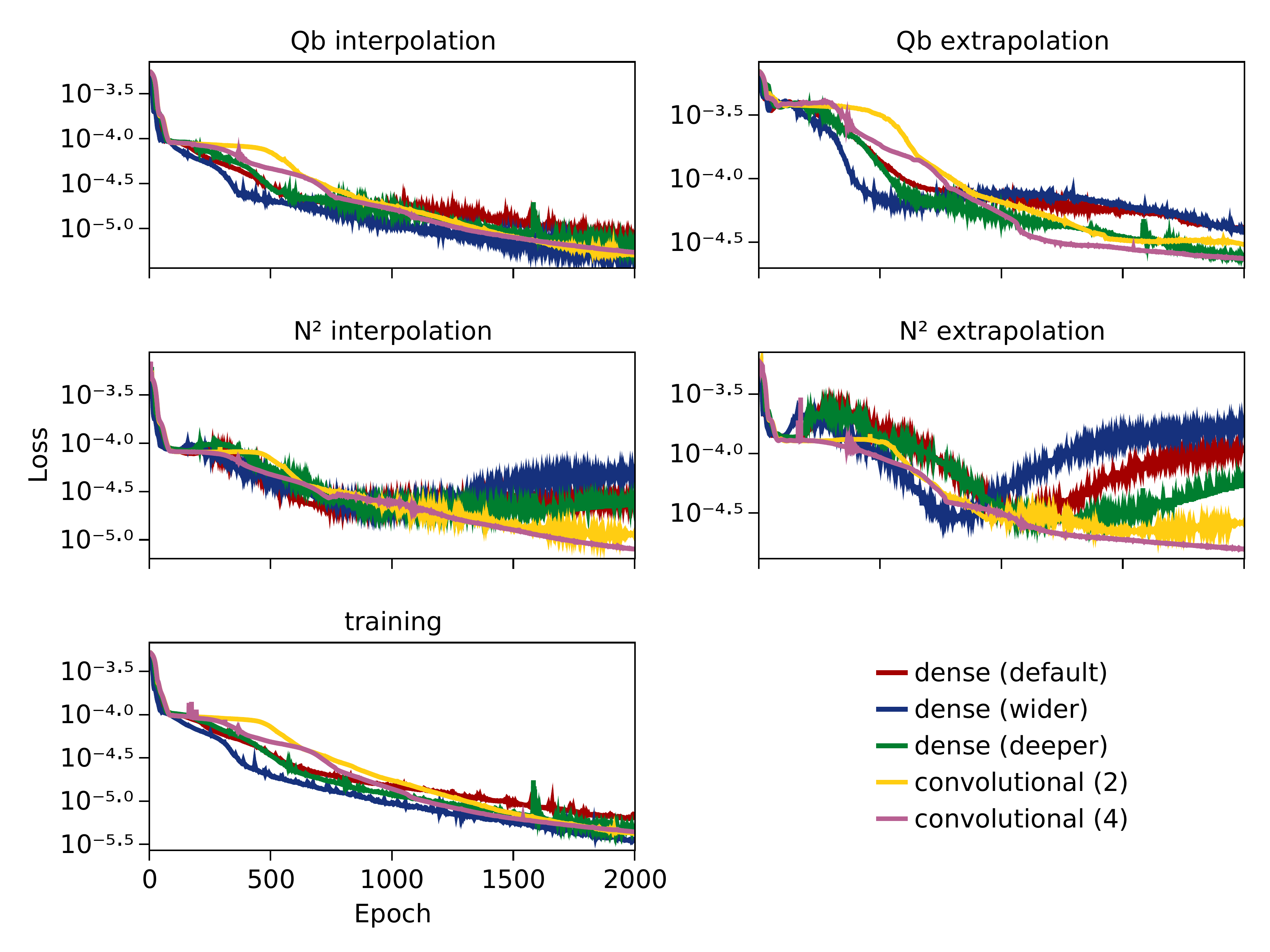}
  	\caption{Time series of loss $\mathcal{L}_2$ [equation \eqref{eq:loss-function-2}] for various neural network architectures for each training and testing simulation, as a function of the training epoch. Each colored curve corresponds to one of the architectures used and each panel shows how the different architectures performed on each set of simulations.}
 	\label{fig:comparing-architectures}
\end{figure}

Figure \ref{fig:comparing-architectures} shows how the different architectures performed on the different sets of simulations. The architectures all perform similarly on the training set. Considering the interpolation and extrapolation capabilities we see that the convolutional (4) architecture consistently outperform the others. In the $N^2$ interpolation and extrapolation cases we see that by epoch 2000 all the architectures are likely overfitting the training data except for convolutional (4). This is perhaps not surprising since free convection does not exhibit non-locality \citep{Souza20} suggesting that only local information is required and that convolutional networks should outperform fully-connected networks. The convolutional (4) architecture also exhibits less high frequency noise in the loss function, perhaps indicating that its gradients are less noisy and/or better behaved leading to a less challanging training process. The convolutional (4) architecture also has the fewest free parameters, suggesting that fully-connected networks perhaps have too many, leading to overfitting of the training date. Perhaps designing the architecture around physical assumptions is more important than designing larger or deeper networks.

An encouraging result is that our NDE approach seems robust to the architecture of the neural network employed in that all five architectures were trainable. This suggesting that the NDE approach could perhaps be used to perform hypothesis testing on the ideal structure of the parameterization. More investigation is necessary to determine whether a single architecture can perform best in all cases. A more systematic and powerful approach to hyperparameter optimization might prove useful here \citep{Snoek12} and specialized inference algorithms for large datasets and networks exist \citep{Klein17}.

\section{Discussion and conclusions} \label{sec:conclusion}

% \begin{enumerate}
%     \item Mention ways to trim the neural network and speed it up.
%     \item some new possibilities, like using dense and convolutional layers to do hypothesis testing on local and nonlocal closures.
%     \item Interpretibility methods in case we want to understand what the neural network is doing.
%     \item Orthogonal polynomial basis layers for resolution-independent NDEs.
%     \item Discuss extension to oceanic convection (momentum and buoyancy fluxes), building up to Xin Kai's work.
% \end{enumerate}

Here we have illustrated how NDEs might provide an attractive medium for developing data-driven parameterizations for climate models. NDEs can be constructed to augment an existing parameterization scheme whilst remaining faithful to conservation laws. They can then be, for example, trained on high-resolution LES data within a differential equation time-stepper. The resulting solutions are stable and faithfully represent turbulent processes which are challenging to capture using conventional schemes. For example, we have shown that a simple parameterization of convection --- that of convective adjustment --- can be improved upon by training the NDE to capture the entrainment fluxes at the base of the well-mixed layer, fluxes that convective adjustment cannot represent. The augmented parameterization outperforms existing commonly used parameterizations such as KPP. Training even stiff NDEs (e.g. when convective adjustment is employed) is made possible by the Julia scientific machine learning (SciML) software stack.

We can usefully describe the method we are advocating as a 'residual' approach, in which the NN is used to improve upon an existing parameterization through the representation of (residual) fluxes which are not captured by the base parameterization. All of this is elegantly enabled through the use of NDEs. We see no reason why a similar residual approach cannot be used to improve upon many other existing parameterizations exploiting the attractive form of NDEs.  

We have chosen to focus on convectively-driven turbulence of the ocean boundary layer since it provides an idealised, proof-of-context setting for our exploration of NDEs. A number of implementation choices were made for simplicity and could be improved. The parameterization was trained and tested using a single vertical resolution. Existing parameterizations such as KPP can be sensitive to model resolution \citep{Souza20} and development of a resolution-independent parameterization would be desirable. One approach might be to decompose the temperature profile into a Fourier or Chebyshev series. The NN could then be trained on polynomial coefficients \citep{Li20}. Profiles sampled at different resolutions might then lead to an NDE that effectively becomes resolution-independent. Going even further, an autoencoder neural network might be used to learn the best representation of the data which could, in principle, be very different from a polynomial expansion, although perhaps less interpretable.

The learning methods and tooling employed here are rather standard in the ML community and we have made no attempt to optimize the NDE's performance for either prediction time or training time. Neural network pruning can be used to eliminate unnecessary free parameters in the architecture leading to faster predictions with minimal impact on accuracy. A more thorough investigation into hyper-parameter optimization could reveal smaller and faster architectures that perform just as well or even better than, for example, those explored in section \ref{sec:hyperparameter-optimization}. Moreover, reducing the amount of training data could lead to similar accuracy yet with shorter training times. In the present study the networks employed were too small to benefit from the speedup provided by GPUs. That said, larger networks might benefit from GPU resources and offer large speedups compared to CPUs. Efficient GPU-enabled NDEs might then be readily embedded into fast GPU-enabled ocean models such as Oceananigans \citep{OceananigansJOSS} and Veros \citep{Hafner18}.

Data-driven parameterizations, residual or not, can also be used to perform hypothesis testing for closures. For example, to test whether a local or nonlocal closure is needed to model some complex physics, two neural networks may be trained: one using dense, fully-connected layers to furnish a fully nonlocal closure and another using convolutional layers for a local closure. Comparing the accuracy of the two could indicate whether a nonlocal closure is required, or provide insight into which physical scenarios exhibit nonlocal physics. There are a vast number of possibilities that could be explored: recurrent neural networks can be used to test whether including time history improves the closure, long-term short-memory (LSTM) architectures and transformers \citep{Vaswani17} to test whether including lagged information improves the closure, and so on. Custom architectures with new physical properties can also readily be constructed, for example, to create a data-driven parameterization that is only nonlocal. Physical insights from such hypothesis testing with data-driven parameterizations may be valuable in and of themselves, and can also guide the development of improved theory-driven parameterizations.

Future work should explore whether the NDE framework can be applied to more complex turbulent processes in both the ocean and atmosphere. Having tackled free convection, it would be natural to consider the more realistic case of a water column forced by both buoyancy and momentum surface fluxes. The parameterization of mesoscale eddy transport in the ocean is also a crucial and inadequately-represented turbulent process for which NDEs might improve upon existing parameterizations. We considered physical scenarios with constant surface fluxes but future work should also investigate whether NDEs are capable of performing well in the presence of spatially- and time-varying surface fluxes. Since surface fluxes are prescribed here as boundary conditions, rather than being fed into the neural network, the NDE approach could generalize rather well to less idealised forcing scenarios.

% Emphasize that the NDE just needs to interpolate well since we can train it on a wide range of physically realistic physical scenarios so that extrapolation does not happen (or happens very rarely).

\begin{appendices}

\section{Numerical methods and Oceananigans.jl} \label{appx:oceananigans}

Oceananigans.jl \citep{OceananigansJOSS} is open source software for ocean studies written in the Julia programming language \citep{Bezanson17}. It runs on both CPUs or GPUs using Julia's native GPU compiler \citep{Besard19}.

For the large eddy simulations performed in this paper, Oceananigans.jl is configured to solve the spatially filtered, incompressible Boussinesq equations with a temperature tracer. Letting $\bm{u} = (u, v, w)$ be the three-dimensional, spatially filtered velocity field, $T$ be the temperature, $p$ be the kinematic pressure, $f$ be the Coriolis parameter, and $\bm{\tau}$ and $\bm{q}$ be the stress tensor and temperature flux due to sub-filter turbulent diffusion, the equations of motion are

\begin{equation} \label{eq:boussinesq_momentum}
    \partial_t \bm{u} + (\bm{u} \cdot \grad) \bm{u} + \bm{f} \times \bm{u} + \grad p = \bm{b} - \grad \cdot \bm{\tau}
\end{equation}

\begin{equation} \label{eq:boussinesq_temperature}
    \partial_t T + \bm{u} \cdot \grad T = - \grad \cdot \bm{q}
\end{equation}

\begin{equation} \label{eq:incompressible}
    \grad \cdot \bm{u} = 0
\end{equation}

The buoyancy $\bm{b} = b \zhat$ appearing in equation \eqref{eq:boussinesq_momentum} is related to temperature by a linear equation of state,

\begin{equation}
    b = \alpha g (T_0 - T)
\end{equation}

where $T_0 = \SI{20}{\celsius}$ is a reference temperature, $\alpha = \SI{2e-4}{\per\kelvin}$ is the thermal expansion coefficient, and $g = \SI{9.81}{\meter\per\second\squared}$ is gravitational acceleration at the Earth's surface.

The sub-filter stress and momentum fluxes are modeled with downgradient closures, such that

\begin{equation} \label{eq:sub_filter_fluxes}
    \tau_{ij} = -2\nu_e\Sigma_{ij} \quad \text{and} \quad \bm{q} = -\kappa_e \grad T
\end{equation}

where $\Sigma_{ij} = (\partial_i u_j + \partial_j u_i) / 2$ is the strain rate tensor and $\nu_e$ and $\kappa_e$ are the eddy viscosity and eddy diffusivity of temperature. The eddy viscosity $\nu_e$ and eddy diffusivity $\kappa_e$ in equation \eqref{eq:sub_filter_fluxes} are modeled with the anisotropic minimum dissipation (AMD) formalism introduced by \citet{Rozema15} and \citet{Abkar16}, refined by \citet{Verstappen18}, and validated and described in detail for ocean-relevant scenarios by \citet{Vreugdenhil18}. AMD is accurate on anisotropic grids \citep{Vreugdenhil18} and is relatively insensitive to resolution \citep{Abkar16}.

To solve equations \eqref{eq:boussinesq_momentum}--\eqref{eq:incompressible} Oceananigans.jl uses a staggered C-grid finite volume spatial discretization \citep{Arakawa77} with an upwind-biased \nth{5}-order weighted essentially non-oscillatory (WENO) advection scheme for momentum and tracers \citep{Shu09}. Diffusion terms are computed using centered \nth{2}-order differences. A pressure projection method is used to ensure the incompressibility of $\bm{u}$ at every time step \citep{Brown01}. A fast Fourier-transform-based eigenfunction expansion of the discrete second-order Poisson operator is used to solve the discrete pressure Poisson equation for the pressure on a regular grid \citep{Schumann88}. An explicit \nth{3}-order Runge-Kutta method is used to advance the solution in time \citep{Le91}.

% And a description of LES which is probably overkill:
% Thanks to advances in computing, large eddy simulations (LES), which approximate the true small-scale dynamics based on the governing, three-dimensional Navier-Stokes Equations, can now be exploited as approximate ``ground truth" data. These large eddy simulations, first conceptualized in 1972 by J.W. Deardroff \cite{deardorff}, are generally touted as the next-most-accurate approach, after direct numerical simulation, for resolving the Navier-Stokes equations—next-most-accurate because they restrict themselves to only resolving the dynamics above a certain length-scale threshold \cite{Large1998}. We accept the sacrifice in accuracy—the sacrifice that comes with leaving the dynamics at the smallest scales unresolved—for the accompanying gains in efficiency. Even still, each simulation typically involves over 10 million degrees of freedom. The simulations are far too slow to be used in full-scale models, which comprise large numbers of individual components that each need to capture long windows of time. They are suitable, however, for this project's purpose, as they are accurate and permit long enough integration times to be useful for training. Simulations are particularly welcome as oceanographers lack the wealth of meteorological observations available to atmospheric scientists that are needed to help uncover deficiencies in parameterizations.

\section{Simulation setup and generation of LES training data} \label{appx:les-setup}

To generate training data we run a suite of high-resolution LES simulations using Oceananigans.jl. The simulations capture various combinations of surface buoyancy flux and stratification profiles keeping everything else constant. Parameter values are chosen so that the mixed layer depth does not extend beyond approximately half of the domain depth by the end of the simulation, thus minimizing significant finite size effects.

In all simulations, the model domain is $\SI{512}{\meter} \times \SI{512}{\meter} \times \SI{256}{\meter}$ with 256 grid points in the horizontal and 128 grid points in the vertical, giving a $\SI{2}{\meter}$ isotropic grid spacing. A Coriolis parameter of $f = \SI{e-4}{\per\second}$ is used. The fluid is initially at rest ($\bm{u} = \bm{0}$) and the initial condition on $T$ is horizontally homogeneous and has three layers. At the top is a \emph{surface} layer of thickness $\Delta z_s$ with constant stratification $N^2_s$. At the bottom is a \emph{deep} layer with constant stratification $N^2_d$. In the middle is a \emph{transition} layer or \emph{thermocline} layer of thickness $\Delta z_t$ with nonlinear stratification determined by fitting a cubic polynomial for $T$ between the surface and deep layers preserving continuity and differentiability. The deep layer then has a thickness of $L_z - \Delta z_s - \Delta z_t$ where $L_z = \SI{256}{\meter}$ is the domain depth. Random noise of the form $\epsilon e^{z/\delta}$ where $\epsilon \sim \mathrm{Uniform}(0, 1)$ and $\delta = \SI{8}{\meter}$ is a noise decay length scale is added to the initial $T$ to stimulate numerical convection near the surface. An example of this initial condition can be seen in figure \ref{fig:training-data}.

The surface buoyancy flux $Q_b$ is implemented as a surface temperature flux boundary condition $Q_\theta = Q_b / (\alpha g)$. A gradient boundary condition on $T$ is imposed at the bottom of the domain to maintain the bottom stratification. A sponge layer of the form $\partial_t \bm{u} = -\tau^{-1} e^{-(z+L_z)/\ell} \bm{u}$, where $\tau = \SI{15}{minutes}$ and $\ell = L_z/10$, is used to relax the velocities to zero near the bottom, in part to absorb internal waves that could otherwise bounce around the domain. The temperature $T$ is similarly relaxed to the initial condition with the form $\partial_t T = -\tau^{-1} e^{-(z+L_z)/\ell} \left[T - T_0(z)\right]$ near the bottom to maintain the bottom stratification where $\tau$ and $\ell$ have the same value as before.

Horizontally-averaged output is written to disk every \SI{10}{minutes} of simulation time. Each simulation is run for \SI{8}{days}.

\section{Simulation parameters} \label{appx:simulation-parameters}

In table \ref{table:simulation-parameters} we tabulate the parameters used to generate training and validation simulation data. Simulations 1--9 were used for training while simulations 10--21 were used for validation. Simulations 1--3 were designed to have a smaller thermocline, while 4--6 have a medium thermocline, and 7--9 have a large thermocline. Simulation 10--12 are used to test for $Q_b$ interpolation, 13--15 for $Q_b$ extraoplation, 16--18 for $N^2$ interpolation, and 19--21 for $N^2$ extrapolation. Figure \ref{fig:parameter-space} shows where in $Q_b-N^2$ parameter space these simulations lie.

\begin{table}[!htb] 
\centering
\begin{tabular}{cSSSSSSS}
    \toprule
    {ID} & {$\Delta z_s$ (\si{\meter})} & {$\Delta z_t$ (\si{\meter})} & {$Q_b$ (\si{\meter\squared\per\second\cubed})} & {$Q_b$ (\si{\watt\per\meter\squared})} & {$N^2_s$ (\si{\per\second\squared})} & {$N^2_d$ (\si{\per\second\squared})} \\
    \midrule
    1  & 48 & 24 & \SI{1e-8}{} & \SI{20.9}{} & \SI{2e-6}{} & \SI{2e-6}{} \\
    2  & 48 & 24 & \SI{3e-8}{} & \SI{62.8}{} & \SI{2e-6}{} & \SI{2e-6}{} \\
    3  & 48 & 24 & \SI{5e-8}{} & \SI{104}{}  & \SI{2e-6}{} & \SI{2e-6}{} \\
    4  & 24 & 48 & \SI{1e-8}{} & \SI{20.9}{} & \SI{1e-6}{} & \SI{2e-6}{} \\
    5  & 24 & 48 & \SI{3e-8}{} & \SI{62.8}{} & \SI{1e-6}{} & \SI{2e-6}{} \\
    6  & 24 & 48 & \SI{5e-8}{} & \SI{104}{}  & \SI{1e-6}{} & \SI{2e-6}{} \\
    7  & 24 & 64 & \SI{1e-8}{} & \SI{20.9}{} & \SI{1e-6}{} & \SI{5e-6}{} \\
    8  & 24 & 64 & \SI{3e-8}{} & \SI{62.8}{} & \SI{1e-6}{} & \SI{5e-6}{} \\
    9  & 24 & 64 & \SI{5e-8}{} & \SI{104}{}  & \SI{1e-6}{} & \SI{5e-6}{} \\
    \midrule
    10 & 48 & 24 & \SI{4e-8}{} & \SI{83.8}{} & \SI{2e-6}{} & \SI{2e-6}{} \\
    11 & 24 & 48 & \SI{2e-8}{} & \SI{41.9}{} & \SI{1e-6}{} & \SI{2e-6}{} \\
    12 & 24 & 64 & \SI{4e-8}{} & \SI{83.8}{} & \SI{1e-6}{} & \SI{5e-6}{} \\
    13 & 48 & 24 & \SI{6e-8}{} & \SI{126}{}  & \SI{2e-6}{} & \SI{2e-6}{} \\
    14 & 24 & 48 & \SI{0.5e-8}{} & \SI{10.5}{} & \SI{1e-6}{} & \SI{2e-6}{} \\
    15 & 24 & 64 & \SI{6e-8}{} & \SI{126}{}  & \SI{1e-6}{} & \SI{5e-6}{} \\
    16 & 36 & 36 & \SI{1e-8}{} & \SI{20.9}{} & \SI{2e-6}{} & \SI{2e-6}{} \\
    17 & 36 & 36 & \SI{5e-8}{} & \SI{104}{}  & \SI{2e-6}{} & \SI{2e-6}{} \\
    18 & 24 & 56 & \SI{3e-8}{} & \SI{62.8}{} & \SI{1e-6}{} & \SI{2e-6}{} \\
    19 & 36 & 36 & \SI{1e-8}{} & \SI{20.9}{} & \SI{2e-6}{} & \SI{2e-6}{} \\
    20 & 36 & 36 & \SI{5e-8}{} & \SI{104}{}  & \SI{2e-6}{} & \SI{2e-6}{} \\
    21 & 24 & 56 & \SI{3e-8}{} & \SI{62.8}{} & \SI{1e-6}{} & \SI{2e-6}{} \\
    \bottomrule
\end{tabular}
\vspace{10pt}
\caption{Simulation parameters used to generate training and validation data. Parameters changed are described in appendix \ref{appx:les-setup}.}
\label{table:simulation-parameters}
\end{table}

\section{Derivation of the non-dimensional PDE} \label{appx:nde-derivation}

The numerical values of $\overline{T} \sim \mathcal{O}(\SI{e1}{\celsius})$ and $\overline{w^\prime T^\prime} \sim \mathcal{O}(\SI{e-5}{\meter\per\second\per\kelvin})$ vary across six orders of magnitude. When training the neural network or performing a gradient descent search, having huge disparities in values may make it difficult to find optimal step sizes and thus difficult to train. Thus we perform a feature scaling on the values of $\overline{T}$ and $\overline{w^\prime T^\prime}$ to normalize the data before processing and training. We use a zero-mean unit-variance scaling 

\begin{equation} \label{eq:T_wT_dimensionless}
    \widehat{\overline{T}} =
    {\frac
        {\overline{T} - \mu_{\overline{T}}}
        {\sigma_{\overline{T}}}
    }
    \quad \mathrm{and} \quad
    \widehat{\overline{w^\prime T^\prime}} =
    \frac
    {\overline{w^\prime T^\prime} - \mu_{\overline{w^\prime T^\prime}}}
    {\sigma_{\overline{w^\prime T^\prime}}}
\end{equation}

where $\widehat{\cdot}$ indicates a normalized quantity and $\mu_\alpha$ and $\sigma_\alpha$ are the mean and standard deviation of $\alpha$ evaluated over the entire training datasets. Using this scaling both the inputs and outputs of the neural network are dimensionless and $\mathcal{O}(1)$.

As the neural network now deals in dimensionless quantities, the NDE being solved, equation \eqref{eq:nde}, must also be non-dimensionalized. This is also important to ensure numerical stability reasons. We non-dimensionalize time $t$ and the vertical coordinate $z$ as

\begin{equation}
    \widehat{t} = \frac{t}{\tau}
    \quad \mathrm{and} \quad
    \widehat{z} = \frac{z}{L_z}
\end{equation}

where $\tau = \SI{8}{days}$ is the duration of the simulation and $L_z = \SI{256}{\meter}$ is the depth of the domain. The time derivative of $\overline{T}$ becomes

\begin{equation} \label{eq:dTdt_dimensionless}
    \p{\overline{T}}{t} = \frac{\sigma_T}{\tau} \p{\widehat{\overline{T}}}{\widehat{t}}
\end{equation}

and the vertical derivative of $\overline{w^\prime T^\prime}$ becomes

\begin{equation} \label{eq:dwTdz_dimensionless}
    \p{\overline{w^\prime T^\prime}}{z}
    =
    \frac
    {\sigma_{\overline{w^\prime T^\prime}}}{L_z}
    \p{\widehat{\overline{w^\prime T^\prime}}}{\widehat{z}}
\end{equation}

Using equations \eqref{eq:T_wT_dimensionless}--\eqref{eq:dwTdz_dimensionless} we can rewrite the NDE, equation \eqref{eq:nde}, non-dimensionally as

\begin{equation} \label{eq:nde_dimensionless}
    \p{\widehat{\overline{T}}}{\widehat{t}}
    =
    - {\frac{\sigma_{\overline{w^\prime T^\prime}}}{\sigma_{\overline{T}}}}
    \left(
        \p{}{\widehat{z}} \widehat{\overline{w^\prime T^\prime}}
        - \widehat{\kappa} \p{\widehat{\overline{T}}}{\widehat{z}}
    \right)
\end{equation}

which is solved numerically during the training process as described in \S\ref{sec:NDEs}.

\section{Calibration of convective adjustment scheme} \label{appx:calibration-ca}

\begin{figure}[!htb]
	\centering
	\includegraphics[width=\linewidth]{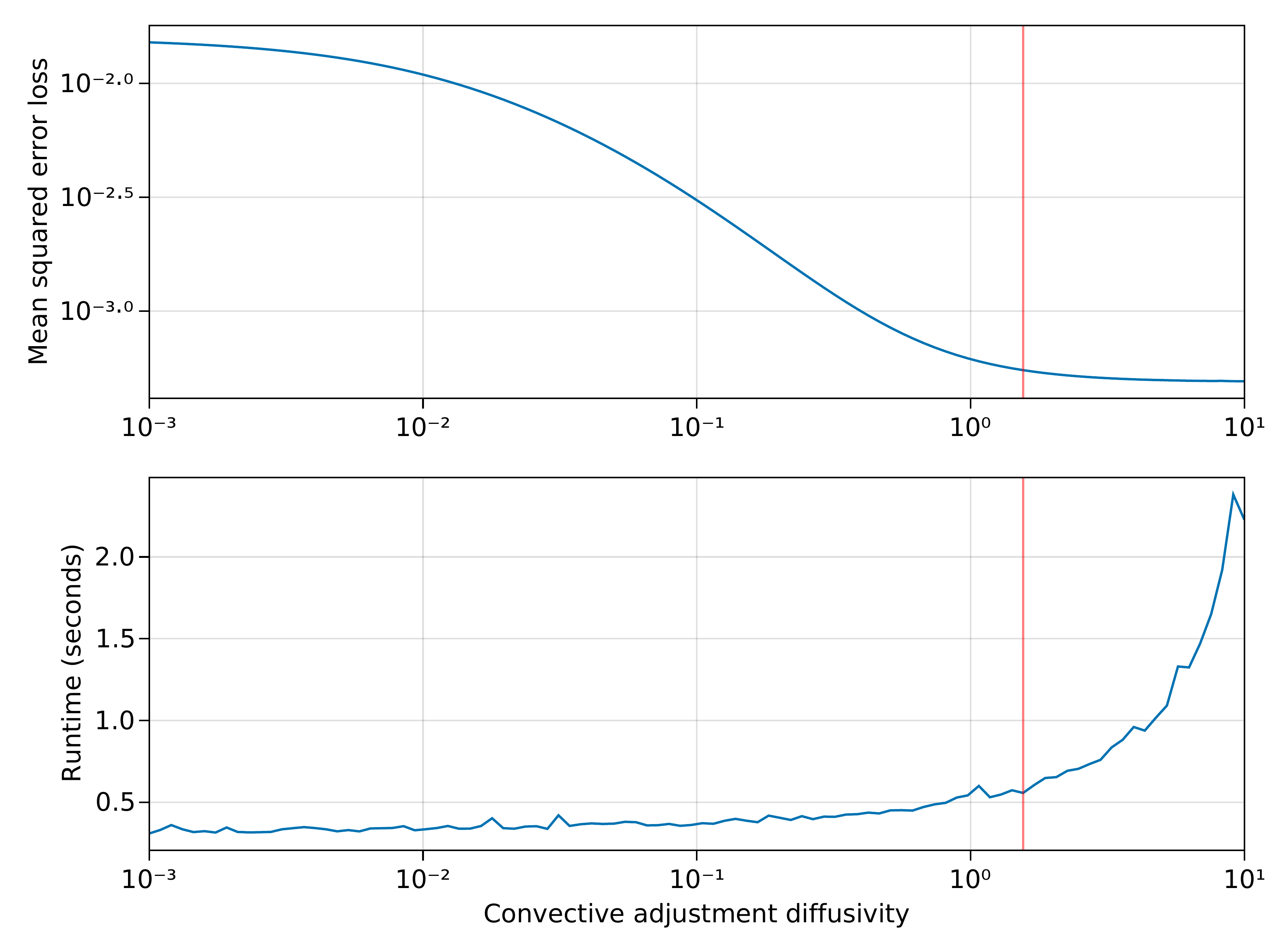}
  	\caption{(Top) The value of the loss function \eqref{eq:loss-function-2} evaluated over the set of training simulations as a function of the convective adjustment diffusivity $K_\mathrm{CA}$. (Bottom) The median wall clock time of the evaluations at each value of $K_\mathrm{CA}$. The red line indicates the value of $K_\mathrm{CA}$ that minimizes the product of the loss and runtime.}
 	\label{fig:calibrating-convective-adjustment}
\end{figure}

Before training the NDE, the base parameterization must be calibrated. In the case of convective adjustment there is only one free parameter, the convective adjustment diffusivity coefficient $K_\mathrm{CA}$ in equation \eqref{eq:convective-adjustment}. Since one can readily run simulations of convective adjustment to compare with LES data we evaluate the loss function \eqref{eq:loss-function-2} over the set of training simulations for values of $10^{-3} \le K_\mathrm{CA} \le 10^1$ to determine the optimal value of $K_\mathrm{CA}$ (see figure \ref{fig:calibrating-convective-adjustment}).

% Mention that this was with adaptive time-stepping?

We see that increasing $K_\mathrm{CA}$ only decreases the loss until it plateaus when $K_\mathrm{CA} > 1$. It would seem natural, then, that we pick a large enough value to readily induce vertical mixing. However, as the NDE implemented in DifferentialEquations.jl uses the ROCK4 explicit time-stepper, increasing $K_\mathrm{CA}$ also increases the stiffness of the system of differential equations necessitating a smaller time step and therefore longtimer runtimes. We therefore choose a value of $K_\mathrm{CA}$ that minimizes the product of the loss function and the runtime. This turns out to correspond to a non-dimensional diffusivity of $\widehat{K}_\mathrm{CA} \approx 2$ or a dimensional value of $K_\mathrm{CA} = (H^2/\tau) \widehat{K}_\mathrm{CA} \approx \SI{0.2}{\meter\squared\per\second}$. This is a very reasonable value and is in the range discussed by \citet{Klinger96}.

\section{Calibration of the K-Profile Parameterization} \label{appx:calibration-kpp}

For the comparison between the NDE and KPP parameterizations described in section \ref{ssec:comparing-parameterizations}, both should be trained on the same set of simulations. This is especially true because KPP may not be optimized for free convection, whilst the NDE is.

We use the KPP implementation provided by OceanTurb.jl and described by \citet{Souza20}. It is a mathematically identical algebraic reorganization of the original formulation proposed by \citet{Large94} to reduce the number of free parameters from six to just four $\bm{\mathcal{C}} = (\mathcal{C}^S, \mathcal{C}^H, \mathcal{C}^D, \mathcal{C}^N)$ in the case of free convection. $\mathcal{C}^S$ is a surface layer fraction, $\mathcal{C}^H$ is a mixing depth parameter, $\mathcal{C}^D$ is a flux scaling parameter for downgradient fluxes, and $\mathcal{C}^N$ is a flux scaling for non-local fluxes. The reference parameters as given by \citet{Large19}, rephrased by the four parameters of \citet{Souza20}, are $\bm{\mathcal{C}}_0 = (0.1, 0.96, 1.36, 6.33)$.

Using reference values as an initial guess, we apply an adaptive differential evolution optimization algorithm \citep{Wang14_differential_evolution} with radius-limited sampling as implemented in BlackBoxOptim.jl \citep{Feldt18} (\texttt{adaptive\_de\_rand\_1\_bin\_radiuslimited} more specifically). This algorithm is suitable for finding global minima in the absence of gradient information from automatic differentiation. In this way we train KPP on our 9 simulations by optimizing our four parameters to minimize the loss function \eqref{eq:loss-function-2}.

The reference parameters lead to a loss of $\mathcal{L}_2(\bm{\mathcal{C}}_0) = \num{1.26e-4}$. The optimization algorithm is given the following box constraints following \citet{Souza20}: $0 \le \mathcal{C}^S \le 1$, $0 \le \mathcal{C}^H \le 8$, $0 \le \mathcal{C}^D \le 8$, $0 \le \mathcal{C}^N \le 8$. After roughly $10,000$ iterations, the differential evolution algorithm converges on the parameters $\bm{\mathcal{C}}^\star = (\frac{2}{3}, 8, 0.16, 5)$ with loss  $\mathcal{L}_2(\bm{\mathcal{C}}^\star) = \num{5.33e-5}$ which corresponds to an improvement of roughly $2.4\times$. $\bm{\mathcal{C}}^\star$ is used when comparing KPP against other parameterizations in section \ref{ssec:comparing-parameterizations}.

\section{Neural network architectures} \label{appx:nn-architectures}

In section \ref{sec:hyperparameter-optimization} we trained the NDE against the same training simulations but using five different architectures, as detailed in table \ref{table:nn-archs}. They are composed of a number of sequential layers, either fully-connected dense layers denoted by $D_{n \rightarrow m}$ with $n$ inputs and $m$ outputs or one-dimensional convolutional layers denoted $C_n$ with filter size $n$. The total number of free parameters for each architecture is included in the table, together with the training time per epoch for both training methods. The size of the dense layers were typically in multiples of $N_z = 32$, the number of vertical grid points.

\newcommand{\archDenseDefault}{$D_{N_z \rightarrow 4N_z}, D_{4N_z \rightarrow 4N_z}, D_{4N_z \rightarrow N_z-1}$}
\newcommand{\archDenseWider}{$D_{N_z \rightarrow 8N_z}, D_{8N_z \rightarrow 8N_z}, D_{8N_z \rightarrow N_z-1}$}
\newcommand{\archDenseDeeper}{$\begin{array} {cc} & D_{N_z \rightarrow 4N_z}, D_{4N_z \rightarrow 4N_z}, \\ & D_{4N_z \rightarrow 4N_z}, D_{4N_z \rightarrow N_z-1} \end{array}$}
\newcommand{\archConvolutionalTwo}{$\begin{array} {cc} & C_2, D_{N_z-2+1 \rightarrow 4N_z}, \\ & D_{4N_z \rightarrow 4N_z}, D_{4N_z \rightarrow N_z-1} \end{array}$}
\newcommand{\archConvolutionalFour}{$\begin{array} {cc} & C_4, D_{N_z-4+1 \rightarrow 4N_z}, \\ & D_{4N_z \rightarrow 4N_z}, D_{4N_z \rightarrow N_z-1} \end{array}$}

\begin{table}[!htb]
\centering
\begin{tabular}{cccSS}
    \toprule
    {Name} & {architecture} & {$N_p$} & {$t_\mathrm{train}^\mathrm{fluxes}$ (\si{\second})} & {$t_\mathrm{train}^\mathrm{timeseries}$ (\si{\second})} \\
    \midrule
    dense (default)   & \archDenseDefault      & 24,735 & 4.89 & 20.76 \\
    dense (wider)     & \archDenseWider        & 82,207 & 6.23 & 35.05 \\
    dense (deeper)    & \archDenseDeeper       & 41,247 & 5.77 & 25.48 \\
    convolutional (2) & \archConvolutionalTwo  & 24,610 & 8.84 & 37.97 \\
    convolutional (4) & \archConvolutionalFour & 24,356 & 8.89 & 38.83 \\
    \bottomrule
\end{tabular}
\vspace{10pt}
\caption{Neural network architectures used in section \ref{sec:hyperparameter-optimization}. $N_p$ is the total number of free parameters for the architecture. $t_\mathrm{train}^\mathrm{fluxes}$ and $t_\mathrm{train}^\mathrm{timeseries}$ are the training times per epoch (in seconds) using loss functions $\mathcal{L}_1$ [equation \eqref{eq:loss-function-1}] and $\mathcal{L}_2$ [equation \eqref{eq:loss-function-2}] respectively.}
\label{table:nn-archs}
\end{table}

% \section{Performance benchmarking}

\end{appendices}

\section*{Open Research}
Julia code to produce the simulation data, train the NDE, and plot all the figures can be found in a GitHub repository (\url{https://github.com/ali-ramadhan/NeuralFreeConvection.jl}) archived at Zenodo \citep{NeuralFreeConvection.jl}.

\section*{Acknowledgements}
We thank Keaton Burns for insightful discussions during the development of this study. Our work is supported by the generosity of Eric and Wendy Schmidt by recommendation of the Schmidt Futures program, by the National Science Foundation under grant AGS-6939393 and by the MIT-GISS collaborative agreement of NASA.

\pagebreak

\bibliography{main}

\end{document}